\documentclass[fleqn,usenatbib,usedcolumn]{mnras}

\usepackage{newtxtext}
\usepackage[slantedGreek]{newtxmath}

\usepackage[T1]{fontenc}
\usepackage{ae,aecompl}

\usepackage{enumerate}
\usepackage{graphicx}	
\usepackage{amsmath}	
\usepackage{amssymb}	
\usepackage{natbib}
\usepackage{longtable,lscape}

\bibpunct{(}{)}{;}{a}{}{,}
\title[The young cluster Be 51]{Berkeley 51, a young open cluster with four yellow supergiants}

\author[I. Negueruela et al.]{
I. Negueruela,$^{1}$\thanks{E-mail: ignacio.negueruela@ua.es}
M. Mongui\'o,$^{1,2}$
A. Marco,$^{1}$
H.~M. Tabernero,$^{1}$\newauthor
C. Gonz\'{a}lez-Fern\'{a}ndez,$^{3}$
and R. Dorda$^{1}$\\
$^{1}$Departamento de F\'{\i}sica, Ingenier\'{\i}a de Sistemas y
Teor\'{\i}a de la Se\~{n}al, Universidad de Alicante, Carretera San Vicente del Raspeig s/n,\\
E03690, San Vicente del Raspeig, Spain\\
$^2$ School of Physics, Astronomy \& Mathematics, University of Hertfordshire, College Lane, Hatfield, Hertfordshire AL10 9AB, UK\\
$^{3}$Institute of Astronomy, University of Cambridge, Madingley Road, Cambridge CB3 0HA, UK\\
}

\date{Accepted XXX. Received YYY; in original form ZZZ}

\pubyear{2018}
   \volume{{\rm in press}}
\begin{document}
\label{firstpage}
\pagerange{\pageref{firstpage}--\pageref{lastpage}}
\maketitle

\begin{abstract}
The heavily obscured open cluster Berkeley~51 shows characteristics typical of young massive clusters, even though the few previous studies have suggested older ages. We combine optical ($UBV$) and 2MASS photometry of the cluster field with multi-object and long-slit optical spectroscopy for a large sample of stars. We apply classical photometric analysis techniques to determine the reddening to the cluster, and then derive cluster parameters via isochrone fitting. We find a large population of B-type stars, with a main sequence turn-off at B3\,V, as well as a large number of supergiants with spectral types ranging from F to M. We use intermediate resolution spectra of the evolved cool stars to derive their stellar parameters and find an essentially solar iron abundance. Under the plausible assumption that our photometry reaches stars still close to the ZAMS, the cluster is located at $d\approx5.5\:$kpc and has an age of $\sim60\:$Ma, though a slightly younger and more distant cluster cannot be ruled out. Despite the apparent good fit of isochrones, evolved stars seem to reside in positions of the CMD far away from the locations where stellar tracks predict Helium burning to occur. Of particular interest is the presence of four yellow supergiants, two on the ascending branch and two others close to or inside the instability strip.

\end{abstract}




\section{Introduction}

Evolved stars in open clusters represent the best test beds for theoretical evolutionary tracks. After the end of Hydrogen burning in their cores, stars evolve towards lower effective temperatures, $T_{{\rm eff}}$, and become, according to their masses, red giants (RGs) or supergiants (RSGs). For a limited range of masses, loops in the HR diagram are expected to bring the stars back to the yellow supergiant region, where they can behave as classical Cepheids \citep[e.g.][]{chiosi}. As an example, in the most recent Geneva tracks \citep{ekstrom12}, stars of solar composition with masses between 5 and~$9\,M_{\sun}$ experience these loops both for zero initial rotation and moderately-high initial rotation, while older isochrones showed this behaviour at higher masses \citep{schaller92}. The exact mass range for which these loops happen depends on the physics of the stellar interior, generally modelled via poorly understood parameters \citep[e.g.][]{chiosi,mf94,salas99,mm00}. In particular, the extent of semi-convection and overshooting, which are not well constrained, have crucial consequences on issues of fundamental importance in our understanding of Galactic chemical evolution, such as the ratio of initial mass to white dwarf mass \citep[e.g.][]{jeff97,weide00} or the boundary between stars that leave white dwarfs as remnants and those that explode as supernovae \citep[SNe; e.g.][]{poelarends08}.

Unfortunately, due to the rarity of high-mass stars and the short duration of the post-H-core-burning phase, most young open clusters are only moderately useful as test beds because of low number statistics \citep[e.g.][]{ekstrom13}. For ages above 100~Ma, on the other hand, the number of RGs increases for a given cluster mass, meaning that several clusters are known to sport large populations of RGs \citep[see][]{mermilliod08}. Finding young open clusters with large populations of evolved stars provides the laboratories that can help constrain the inputs of models and hence our understanding of stellar evolution \citep{neg16iaufm}.

As part of such an endeavour, we have been searching through the databases of poorly-studied known open clusters to identify good candidates to massive young open clusters. Recent examples include the starburst cluster vandenBergh-Hagen~222, with a population of 13 yellow or red supergiants at an age $\sim16$\,--\,20~Ma \citep{marco14}, or the $\sim$50~Ma open cluster Berkeley~55, with 6\,--\,7 supergiants or bright giants \citep{negmar12}. Here we report on the identification of another faint, Northern sky cluster as a young, massive cluster with a large population of evolved stars.

Berkeley~51 (Be~51) is a faint, compact cluster in the constellation Cygnus. The WEBDA database\footnote{At {\tt http://webda.physics.muni.cz}} \citep{webda} provides coordinates RA:~20h 11m 54s, Dec:~$+34^{\circ}\:24\arcmin\:06\arcsec$ ($\ell=72\fdg15$, $b=+0\fdg29$). Two previous estimates by \citet{tadross} and \citet{kharchenko13}, based on existing photometric catalogues, agree on considering Be~51 an intermediate-age, distant cluster ($\tau=150$~Ma, $d=3.2$~kpc in \citealt{tadross}; $\tau=180$~Ma, $d=3.3$~kpc in \citealt{kharchenko13}, who estimate a reddening $A_{V}=5.8$~mag). In contrast, a $BVI$ CCD study by \citet{subram10} concluded that it was an old ($\tau=1$~Gyr) cluster at only $d=1.3$~kpc, despite its high reddening of $E(B-V)=1.6$. In this paper, we present a much more complete study of Be~51, including comprehensive spectroscopy, that reveals it as a distant, moderately-massive young open cluster containing a large population of evolved stars.

\section{Observations}

\subsection{Optical photometry}

 $UBV$ photometry of Be~51 was obtained in service mode using ALFOSC on the Nordic Optical Telescope at the Roque de los Muchachos Observatory (La Palma, Spain) on the night of 2008 September 19. In imaging mode, the camera covers a field of $6\farcm5 \times 6\farcm5$ and has a pixel scale of $0\farcs19$/pixel.

Two standard fields from the list of \citet{landolt92}, MARK\_A and PG~2213$-$006, were observed during the night in order to provide standard stars for the transformation. Since there is only one measurement of each field, we could not trace the extinction during the night, and so we used the median extinction coefficients for the observatory. The images were processed for bias and flat-fielding corrections with the standard procedures using the CCDPROC package in IRAF\footnote{IRAF is distributed by the National Optical Astronomy Observatories, which are operated by the Association of Universities for Research in Astronomy, Inc., under cooperative agreement with the National Science Foundation}. Aperture photometry using the PHOT package inside DAOPHOT (IRAF, DAOPHOT) was developed on these fields with the same aperture, 21 pixels, for each filter.

Images of Be~51 were taken in two series of different exposure times to obtain accurate photometry for a magnitude range. The log of observations is presented in Table~\ref{photlog}. Photometry was obtained by point-spread function (PSF) fitting using the DAOPHOT package
\citep{stetson87} provided by IRAF. The apertures used are of the order of the FWHM, 5 pixels for all images in the $U$ and $B$ filters and 4 pixels for the $V$ filter images. We selected $\approx 20$ PSF stars in each frame, from which we determined an initial PSF, which we allowed to be variable (in order 2) across the frame. We then performed aperture correction to obtain instrumental magnitudes for all stars. Using
the standard stars and the median extinction coefficients for the observatory, we carried out the transformation of the instrumental magnitudes to the standard system by means of the PHOTCAL package inside IRAF. 

\begin{table}
\caption{Log of the photometric observations taken at the NOT on June 2010 for Berkeley 51. There are two observations for each exposure time.\label{photlog}}
\centering
\begin{tabular}{c c c}
\hline\hline
&\multicolumn{2}{c}{Exposure times (s)}\\
Filter & Long times & Short times \\
\hline
$U$ & 900 & 250\\
$B$ & 200 & 60 \\
$V$ & 40& 10 \\
\hline
\end{tabular}
\end{table}

\begin{figure}
\resizebox{\columnwidth}{!}{\includegraphics{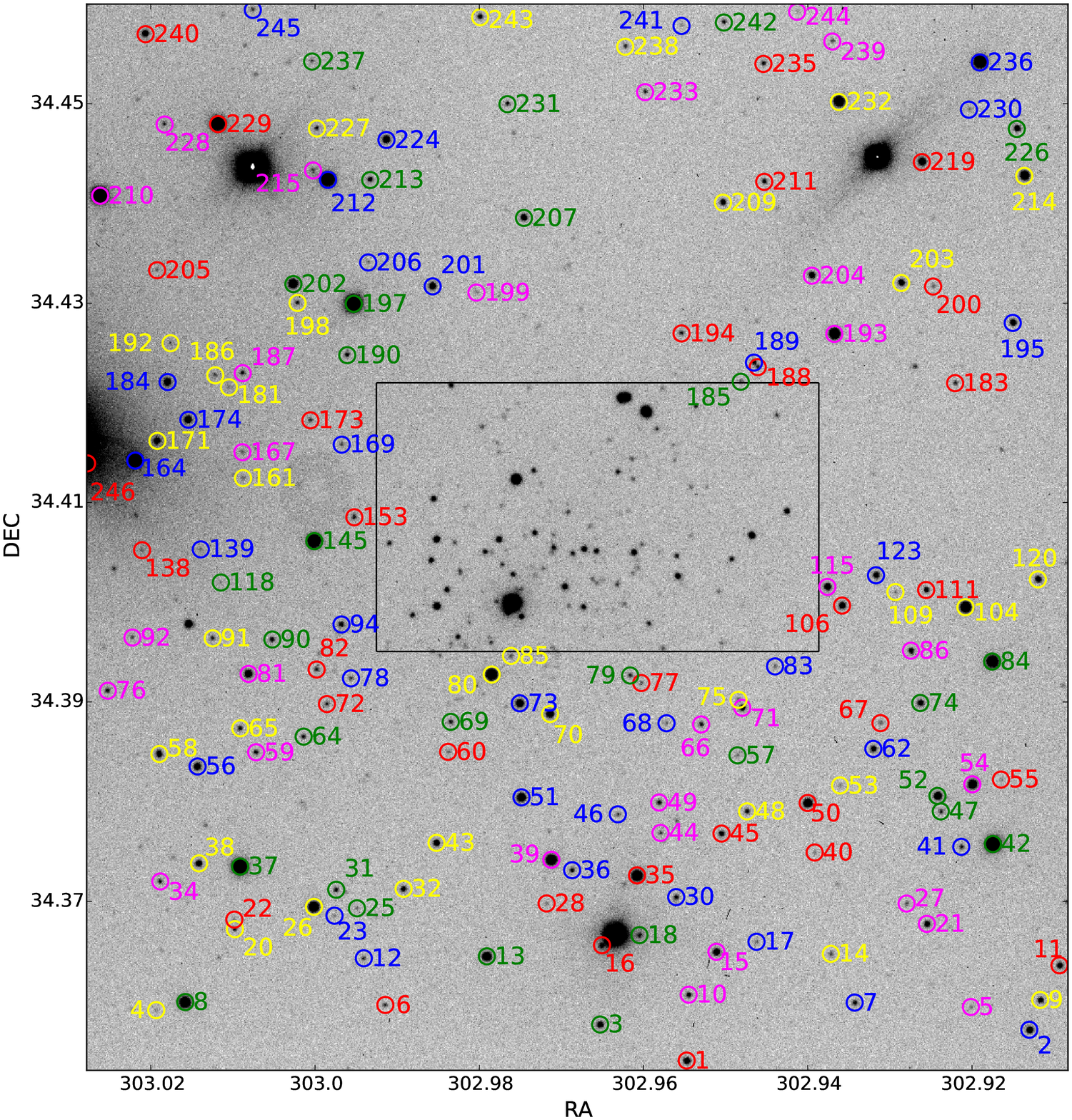}}
\centering
\caption{Finding chart for stars with photometry in the field of Berkeley 51. The image is one of our deep $U$-band frames. Stars inside the rectangle (which approximately defines the cluster core) are marked in Fig.~\ref{findercentre}. Each star is identified by the nearest marker in the same colour as the circle around it (colours are assigned simply for visibility). The size of the image is the full FoV of ALFOSC. \label{findermain} } 
\end{figure}

\begin{figure*}
\resizebox{\textwidth}{!}{\includegraphics{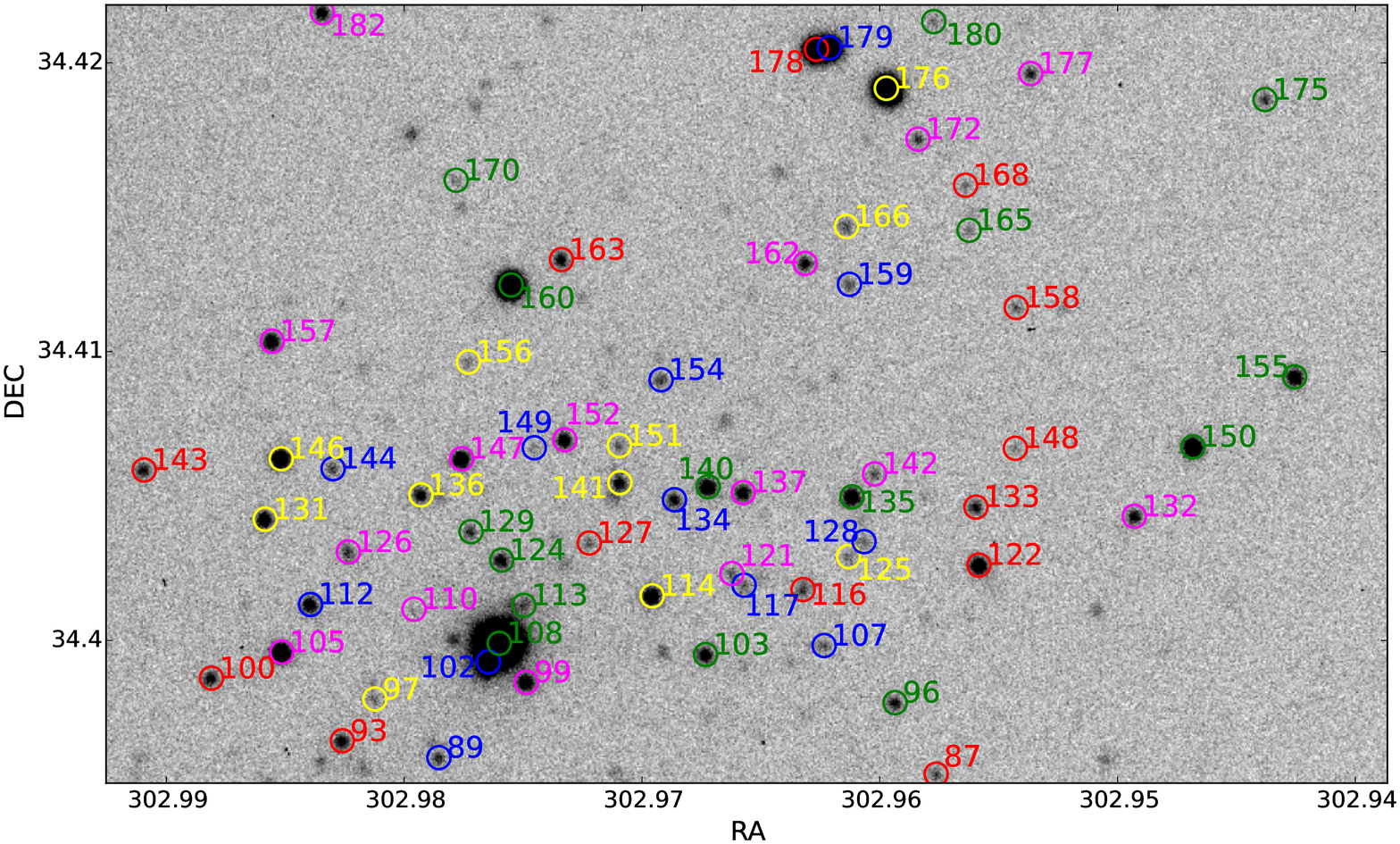}}
\centering
\caption{Finding chart for stars with photometry in the central part of Berkeley 51. Labelling as in Fig.~\ref{findermain}. \label{findercentre}} 
\end{figure*}

The number of stars that we could detect in all filters is limited by the long exposure time in the $U$ filter. We identify all stars with good photometry in all three filters on the image in Figures~\ref{findermain} and~\ref{findercentre}. In Table~\ref{allphot}, we list coordinates (obtained by a cross-match with 2MASS), their $UBV$ photometry, and their 2MASS $JHK_{\textrm{S}}$ photometry, when available. The values of $V$, $(B-V)$ and $(U-B)$ are given together with the number of measurements and an error, which is the standard deviation of all the measurements whenever several measurements exist and the photometric error otherwise. The designation of each star is given by the number indicated on the images (Figures~\ref{findermain} and~\ref{findercentre}). We have three-band photometry for $\sim$250 stars in the field, but in the analysis we will only use 173 stars with photometric errors such that the error in their reddening-free $Q$ parameter (see Sect.~\ref{hr}) is $<0.07$~mag (roughly corresponding to an uncertainty of one spectral type in photometric classification). 

\subsection{2MASS data}
\label{2mass}

We obtained $JHK_{{\rm S}}$ photometry from the 2MASS catalogue \citep{skru06}. The completeness limit of this catalogue is set at $K_{{\rm S}}=14.2$. We selected only stars with ``good'' quality flags in 2MASS (A or E), and photometric errors $<0.05$~mag in all bands. This leaves out many stars close to the centre of the cluster, where confusion becomes important at the spatial resolution of 2MASS.

\begin{figure}
   \centering
\resizebox{\columnwidth}{!}{\includegraphics[clip]{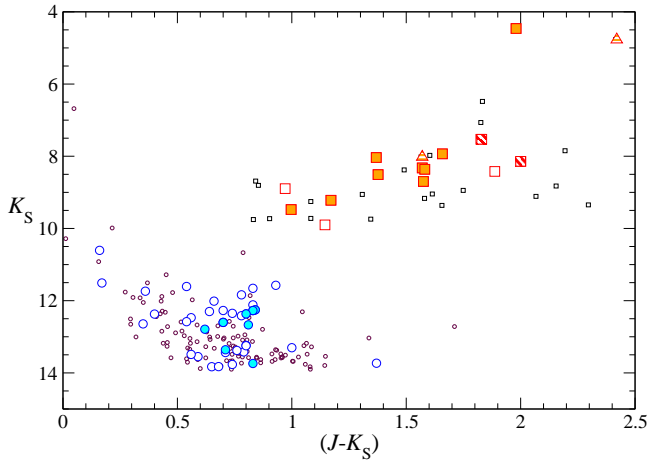}}
   \caption{Colour-magnitude diagram for 2MASS data in circles of radius $3\arcmin$ (large symbols) and $7\arcmin$ (small symbols) around the position of Be~51. Circles represent objects selected as possible early-type stars, while squares are candidate luminous stars (selected as described in the main text). Filled symbols represent stars with spectra classified as likely members. Striped squares are stars with spectra that we do not consider as members. The triangles are more distant objects whose connection to the cluster is unclear. \label{fig:raw}}
    \end{figure}

We used the 2MASS data to carry out a preliminary selection of spectroscopic targets for an exploratory survey. We took a circle of radius $3\arcmin$ around the nominal cluster centre and built the corresponding $K_{{\rm S}}/(J-K_{{\rm S}})$ diagram (see Fig.~\ref{fig:raw}). We cleaned the diagram by making use of the reddening-free $Q_{{\rm IR}}$ index, defined as $Q_{{\rm IR}} = (J-H)-1.8\times(H-K_{\textrm{S}})$. Early-type (OBA) stars are easily separated, as they display $Q_{{\rm IR}}\approx0.0$ \citep[cf.][]{cp05,ns07}. We selected stars with $-0.15\leq Q_{{\rm IR}}\leq0.08$ (shown as large circles in Fig.~\ref{fig:raw}). This range is intended to account for the typical errors in 2MASS (generally $0.03-0.05$~mag in a given colour for stars with $K_{{\rm S}}=12-13$) and also include emission-line stars, which typically have $Q_{{\rm IR}}\la-0.05$ \citep[e.g.][]{ns07}.

In addition, we selected bright stars with $K_{{\rm S}}\leq10.0$ and $0.1\leq Q_{{\rm IR}}\leq0.4$, the range where Galactic red luminous stars are generally found \citep[see][for a discussion of these criteria]{negueruela12, carlos15}. These objects are displayed as big squares in Fig.~\ref{fig:raw}. We then aplied the same criteria to a circle of radius $7\arcmin$ around the nominal cluster centre (displayed as small symbols in Fig.~\ref{fig:raw}). Comparison of  both datasets shows a clear overdensity of bright stars in the central cluster area (almost half the bright stars in the large circle are inside the small circle, which has less than one fifth of the area). In addition, there is a strong overdensity of early type stars with $K_{{\rm S}}=11-13$ in the central area, with most of them presenting $(J-K_{{\rm S}})\approx0.75$. We interpreted this overdensity as the cluster main sequence, and selected targets for the spectroscopy runs among these objects.

\label{sec:maths} 

\subsection{Spectroscopy}
Spectroscopy of the brightest candidate members of Be~51 was obtained with the red arm of the ISIS double-beam spectrograph, mounted on the 4.2-m William Herschel Telescope (WHT) in La Palma (Spain) in three separate runs. An initial survey was conducted with the R600R grating and  the {\it Red+} CCD, a configuration that covers the 7600\,--\,9000\,\AA\ range in the unvignetted section of the CCD with a nominal dispersion of 0.5\,\AA/pixel. Data were taken during a service night on 2007, July 26  and then completed during a run on 2007, August 21. In July, the CCD was unbinned, and a $1\farcs5$ slit was used. In August, the CCD was binned by a factor 2 in the spectral direction, and a  $1\farcs2$ slit was used. For this grating and all slit widths $>1\farcs1$, the spectral resolution is oversampled, and the resolution element is expected to be $\sim$4~unbinned pixels. This has been checked by measuring the width of arc lines, which is on average $\approx2.1\,$\AA\ for both configurations. The resolving power of our spectra is therefore $R\sim4\,200$.

The supergiants identified during this survey were then re-observed at higher resolution, using the R1200R grating in June 2012 and the unbinned {\it Red+} CCD. The nominal dispersion is 0.26\,\AA/pixel. We used a $0\farcs9$ slit that provides a resolving power of $R\sim12\,000$. Unlike in the previous runs, each spectrum was taken together with an arc exposure at the same sky position, for accurate wavelength calibration. A log of all the WHT/ISIS observations is presented in Table~\ref{log}. The average number of counts per pixel in the continuum around the Ca\,{\sc ii} triplet is given to estimate the signal-to-noise (S/N) ratio.

\begin{table}
 \centering
  \caption{Log of the WHT observations. The upper panel shows observations of cool luminous stars confirmed as members, all of which have been observed at least twice. The middle panel contains other cool luminous stars. The bottom panel includes candidate blue stars, observed only once.\label{log}}
  \begin{tabular}{lccc}
\hline
\hline
 \noalign{\smallskip}
   Star   & Exposure &Date & Counts/pixel$^{a}$    \\
& Time (s) &  (UT) & \\
 \noalign{\smallskip}
 \hline
 \noalign{\smallskip}
126  &   30  & 21 Aug 2007 & 35\,000\\
    &   350 & 28 Jun 2012 & 45\,000\\ 
    &   200 & 2 Jul 2012 & 30\,000\\ 
70 &   200 & 26 Jul 2007 & 24\,000\\
    &   450 & 28 Jun 2012 & 10\,000\\ 
134 &   200 & 26 Jul 2007 & 11\,000\\ 
    &   400 & 2 Jul 2012 & 3\,000\\ 
301 &   200 & 26 Jul 2007 & 15\,000\\ 
    &  500  & 28 Jun 2012 & 6\,000\\ 
172 &   200 & 26 Jul 2007 &  12\,000\\
    &   350 & 2 Jul 2012 & 3\,000\\ 
162 &   200 & 26 Jul 2007 &  14\,000\\   
    &  600  & 2 Jul 2012 & 9\,000\\ 
302&   200 & 26 Jul 2007 &  9\,000\\ 
    &   350 & 2 Jul 2012 & 2\,000\\ 
146 &   200 & 21 Aug 2007 &  16\,000\\
    &   350 & 2 Jul 2012 &  3\,000\\
105&   200 & 21 Aug 2007 &  15\,000\\
    &  450  & 28 Jun 2012 & 4\,000\\
\noalign{\smallskip}
\hline
\noalign{\smallskip}
501& 200 &  22 Aug 2007 &18\,000 \\
502& 400 &  22 Aug 2007 & 12\,000\\ 
503& 200 & 22 Aug 2007 &15\,000\\
    & 350   & 2 Jul 2012 &4\,500\\ 
901 &  200 &  2 Jul 2012 &9\,000 \\ 
\noalign{\smallskip}
\hline
\noalign{\smallskip}
114& 900 & 21 Aug 2007 & 7\,500\\
103& 1200&  22 Aug 2007 & 4\,000\\
122& 900 &  21 Aug 2007 & 6\,000\\
150& 1200 & 22 Aug 2007 & 13\,000\\
143& 900 &  21 Aug 2007 & 7\,000\\
147& 900 &  21 Aug 2007 & 7\,000\\
\noalign{\smallskip}
\hline
\end{tabular}
	\begin{list}{}{}
\item[]$^{a}$ Counts per pixel in the spectral direction after extraction of the whole slit width. Note that spectra from August 2007 were pre-binned by a factor of 2 to match the resolution element. For a correct estimation of the SNR, all the other spectra should have two pixels added.
\end{list}
\end{table}

Candidate blue stars in Be~51, selected from our $UBV$ photometry, were observed on the night of 2014 August 24th with the Optical System for Imaging and low-Intermediate-Resolution Integrated Spectroscopy (OSIRIS) instrument, mounted on the 10.4-m Gran Telescopio Canarias (GTC) in La Palma (Spain). The instrument operated in the MOS mode, with $1\farcs2$ slitlets traced on a plate. The R2000B grism covers the nominal range 3950\,--\,5700\,\AA, but the actual spectral coverage for a given object is a strong function of its position on the plate. The nominal dispersion of the R2000B grism is 0.9\,\AA\ per binned pixel (the standard Marconi CCD42-82 mosaic is used in $2\times2$ binned mode). The resolving power, measured on arc lamp spectra, is $R\approx1\,400$.

Only one plate was observed, as the effective field of $7\farcm6 \times 6\farcm0$ is much larger than the cluster size. Three 1895~s exposures were obtained. We reduced these data using the \textit{Starlink} \citep{currie14} software packages CCDPACK \citep{draper} and FIGARO \citep{shortridge} by following standard procedures. The spectra have been normalized to the continuum using DIPSO \citep{howarth14}.

   \begin{figure}
   \centering
\resizebox{\columnwidth}{!}{\includegraphics[angle=-90,clip]{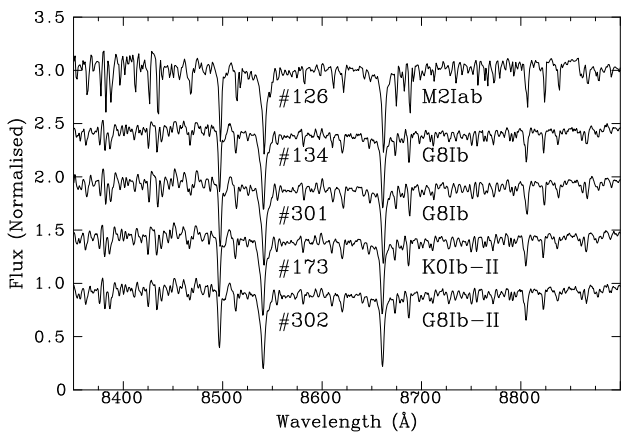}}
   \caption{Intermediate-resolution spectra, taken with ISIS, in the region around the \ion{Ca}{ii} triplet of five bright red stars in the central concentration of Be~51.\label{rsgs}}
    \end{figure}

Finally, we observed one star (\#126) with the High Efficiency and Resolution Mercator Echelle Spectrograph (HERMES), operated at the 1.2~m Mercator Telescope (La Palma, Spain) on 2011 June 15th.  HERMES reaches a resolving power $R= 85\,000$, and a spectral coverage from 377 to 900~nm, though some small gaps exist beyond 800~nm \citep{raskin14}. Data were homogeneously reduced using version 4.0 of the HermesDRS\footnote{http://www.mercator.iac.es/instruments/hermes/hermesdrs.php} automated data reduction pipeline, which provides order merged spectra. In this case, the target is very faint and there are essentially no counts shortwards of $\sim 5\,500\,$\AA.

\section{Results}

\subsection{Spectral classification}

Figure~\ref{rsgs} shows the spectra of five bright red stars in the field preselected as possible evolved members. Classification criteria for spectra at this resolution are discussed in \citet{negueruela12}. The brightest star in the infrared, \#126, is a luminous supergiant of spectral type M2\,Iab. The four other stars are very similar to each other. They are all low-luminosity supergiants with spectral types close to K0 (see Table~\ref{partab}). We have at least two spectra for each star, and in all cases the spectra are consistent with the same classification.

  \begin{figure}
   \centering
\resizebox{\columnwidth}{!}{\includegraphics[angle=-90,clip]{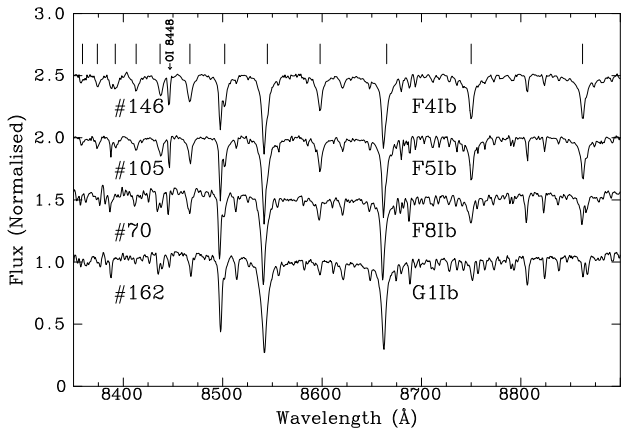}}
   \caption{Intermediate-resolution spectra, taken with ISIS, in the region around the \ion{Ca}{ii} triplet of four bright yellow stars in the central concentration of Be~51. The vertical dashes indicate the positions of the Paschen lines, which weaken as we move along the spectral sequence (note that Pa~13, 15 \& 16 blend with the stronger lines of the triplet). The \ion{O}{i}~8448\,AA\ line, also marked, shows a similar behaviour.\label{ysgs}}
    \end{figure}

Figure~\ref{ysgs} shows the spectra of four candidate luminous stars with fainter $K_{\textrm{S}}$ magnitudes and bluer colours. Comparison to MK standards observed at similar resolution \citep[e.g.][]{cenarro01} suggests that they are all F-type supergiants. This can be confirmed by measuring the strength of their \ion{Ca}{ii} triplet lines \citep[e.g.][]{diaz89, mallik97}. Two of the stars, \#146 and \#105, have very similar spectral types, even though they appear somewhat different because of different rotational velocity. They are mid-F supergiants. The other two stars present clear differences between the two spectra taken at different epochs. Star 70 is a late F supergiant. Both spectra are close to F8\,Ib, but the 2012 one looks discernibly earlier. In the case of \#162, the differences are larger. It is F8\,Ib in the 2012 spectrum, but clearly later (around G1\,Ib) in the 2007 spectrum. As we will see in the following sections, both stars lie on positions compatible with the blue loop in the photometric diagrams\footnote{In a forthcoming paper, Lohr et al.\ (submitted to A\&A) show that \#162 is a 10-d Cepheid.}. The spectral types of all the cool stars are listed in Table~\ref{partab}. Note that two objects lack optical photometry and have been labelled as stars 301 (2MASS~J20115344+3424427) and 302 (2MASS~J20114858+3424420).

\begin{figure}
   \centering
\resizebox{\columnwidth}{!}{\includegraphics[angle=-90,clip]{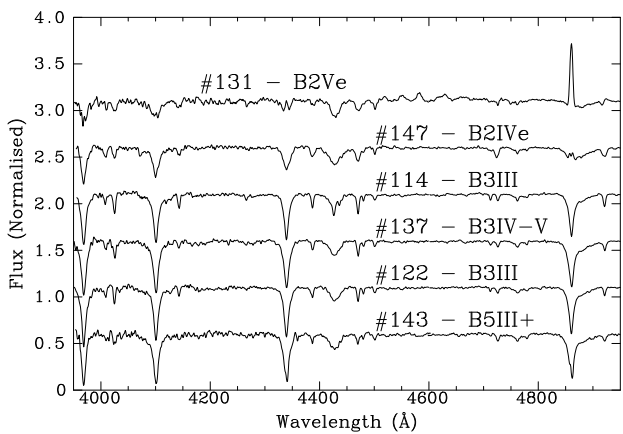}}
   \caption{GTC classification spectra of B-type stars in Be~51.\label{blues}}
    \end{figure}

The GTC/OSIRIS observation provided classification spectra for ten stars, which are listed in Table~\ref{intrinsic_members}. Figure~\ref{blues} shows six of them. The S/N ratio varies greatly among them, but all allow spectral classification. We performed the classification via comparison to spectra of MK standard stars degraded to the same resolution. All the stars observed (corresponding to the top of the cluster blue sequence) are early to mid B stars. Due to high-reddening, the S/N is quite low around H$\gamma$ and, when possible, we have used the set of metallic and \ion{He}{i} lines in the 4\,900\,--\,5100\,\AA\ range to improve the luminosity classification. The spectral types derived are listed in Table~\ref{intrinsic_members}. The spectrum of star 143 shows very asymmetric Balmer lines, most likely indicating the presence of two stars (Fig.~\ref{blues}), one with spectral type $\sim$B5\,III, and an earlier, less bright companion. 
The spectrum of \#153 (not shown) is quite poor and seems to be blended with a late-type interloper (perhaps simply due to bad sky subtraction), but a $\sim$B3\,V classification is likely. 

All the intrinsically blue stars that were observed in the $Z$-band with the WHT were re-observed with GTC in the classification region, except for \#103. For this object, we can derive an approximate spectral type B2\,V from the WHT spectrum. This spectral classification is, however, much less accurate than those based on the blue spectral region and this star will not be used to compare spectroscopic and photometric characteristics. One of the brightest candidates based on the 2MASS colours, star 150, turns out to be a low-luminosity A-type star, and therefore a foreground object.

\subsection{Spectroscopic analysis}

We used the higher resolution ISIS spectra of the cool stars to compute effective temperature ($T_{\textrm{eff}}$) and iron abundance by comparing them to a previously generated grid of synthetic spectra. We employed the new version of the automated code {\scshape StePar} \citep[see][]{tab18}, which relies upon spectral synthesis instead of equivalent widths (EWs) and uses a Markov-chain Montecarlo algorithm \citep[emcee, see][]{fma13} for optimisation. We explored the parameter space using 20 Markov-Chains of 1\,250 points starting from an arbitrary point. As objective function we used a $\chi$-squared in order to fit any previously selected spectral features.

The synthetic spectra were generated using two sets of one-dimensional LTE atmospheric models, based on $15\:$M$_{\sun}$ and $1\:$M$_{\sun}$ MARCS spherical atmospheric models \citep{gus08}. The radiative transfer code employed was \textit{spectrum} \citep{graco94}. As line list, we employed a selection from the VALD database \citep{pis95,kup00,rya15}, taking into account all the relevant atomic and molecular features (dominated by TiO and CN) that can appear in cool luminous stars up to mid-M types. In addition, as Van der Waals damping prescription we employed the Anstee, Barklem, and O'Mara theory (ABO), when available in VALD \citep[see][]{bar00}. Effective temperature $T_{\rm eff}$ ranges from $4\,000\:$K to $8\,000\:$K with a step of $250\:$K for the spectra generated using $1\:$M$_{\sun}$ atmospheric models. For the $15\:$M$_{\sun}$ MARCS synthetic models, $T_{\textrm{eff}}$ varies from $3\,300\:$K to $4\,500\:$K; the step is $250\:$K above $4\,000\:$K and $100\:$K otherwise. Finally, the metalicity ranges from [M/H]$=-1.0\:$dex to [M/H]$=1.0\:$dex in $0.25\:$dex steps for $1\:$M$_{\sun}$ models, whereas $15\:$M$_{\sun}$ MARCS models cover only from [M/H]$=-1.0\:$dex to [M/H]$=0.5\:$dex, in $0.25\:$dex increments. Surface gravity ($\log\,g$) varies from $-0.5$ to 2.0~dex in 0.5~dex steps, when available in each MARCS grid. We convolved our grid of synthetic spectra with a gaussian kernel (FWHM$\approx30\:$km\,s$^{-1}$) to account for the instrumental broadening.

The present version of {\scshape StePar} allows the derivations of any set of stellar atmospheric parameters simultaneously. In this case, we restricted them to only two variables, $T_{\textrm{eff}}$ and metallicity ([M/H]). Given the strong degeneracy between $\log\,g$ and metallicity, surface gravity was kept fixed to values compatible with the position of stars in the observational HR diagram (see Sect.~\ref{params} and compare to Table~\ref{partab}); microturbulence ($\xi$) was adjusted according to the 3D model based calibration described in \citet{dut16}. Since our stars are more luminous than any of their calibration benchmarks we have checked the consistency of our results by obtaining parameters with $\xi$ fixed to $3\:\mathrm{km}\,\mathrm{s}^{-1}$ and $5\:\mathrm{km}\,\mathrm{s}^{-1}$ (i.e.\ higher than any value used) for all stars. The derived $T_{\mathrm{eff}}$'s are quite similar in all cases. Increasing $\xi$ has some effect on the metallicity derived (higher $\xi$ implies lower [M/H]), but even for the extreme case it is less than $\sim0.2$~dex lower.  Our analysis employs some empirically selected lines of Mg, Si, Ti, and Fe \citep{dor16b} in the spectral range around the CaT, $8\,400$\,--\,$8\,900\:$\AA{}. The results of the analysis are listed in Table~\ref{partab}.

\begin{table*}
        \centering
        \caption{Stellar parameters for cool stars in the field of Berkeley~51. The top panel lists stars in the central condensation that are considered likely members. The bottom panel presents two other stars at higher distances.}
        \label{partab}
        \begin{tabular}{lccccccccc}
                \hline
                \noalign{\smallskip}
Star &  Spectral &Date & $T_{\textrm{eff}}$ & $\log\,g$ &  [M/H] & $\xi$ & MARCS grid  & $v_{{\rm hel}}$& $v_{{\rm LSR}}$\\ 
& type & & (K) & & (dex) & (km~s$^{-1}$)& & (km~s$^{-1}$) & (km~s$^{-1}$) \\
\noalign{\smallskip}
\hline
\hline
\noalign{\smallskip}
126  & M2\,Iab & 2012 Jun 28 & $3666\pm39$  & 0.0 & $0.23\pm0.05$ & 1.97 & $15\:$M$_{\sun}$ & $-13$ & $-8$\\
126 & & 2012 Jul 2 & $3656\pm40$  & 0.0 & $0.23\pm0.05$ & 1.96 & $15\:$M$_{\sun}$& $-10$ &$-5$\\
70  & $\sim$F8\,Ib &2012 Jun 28 &--- & --- & --- & ---  & --- &$-26$ & $-21$\\ 
134  & G8\,Ib & 2012 Jul 2 & $4502\pm102$ & 1.0 & $0.04\pm0.06$ & 2.07 & $1\:$M$_{\sun}$& $-7$ & $-2$\\ 
301  & G8\,Ib & 2012 Jun 28 & $4560\pm75$ & 1.0 & $0.31\pm0.04$ & 2.14 & $1\:$M$_{\sun}$&$-11$ & $-6$\\ 
172  & K0\,Ib-II & 2012 Jul 2 & $4537\pm95$ & 1.0 &  $0.16\pm0.06$ & 2.11 & $1\:$M$_{\sun}$&$-24$ & $-19$\\ 
162  & F8\,Ib$^{\dagger}$ & 2012 Jul 2 & $6575\pm65$ & 1.5 & $0.14\pm0.03$ & 3.77 & $1\:$M$_{\sun}$&$-11$&$-16$\\ 
302  & G8\,Ib-II &2012 Jul 2 & $4432\pm130$ & 1.0 & $0.16\pm0.06$ & 1.99 & $1\:$M$_{\sun}$&$-8$&$-3$\\
146  & F4\,Ib & 2012 Jul 2 & ---  & --- &   ---  & --- & ---  & $-4$ &$+1$   \\
105  & F5\,Ib & 2012 Jun 28 & $6727\pm103$ & 2.0 & $0.14\pm0.07$ & 3.25 & $1\:$M$_{\sun}$& $-9$ &$-4$\\
\noalign{\smallskip}
\hline
\noalign{\smallskip}
503 & K2\,Ib &2012 Jul 2 & $4275\pm97$ & 1.0 & $0.16\pm0.05$ & 1.8 & $1\:$M$_{\sun}$& $-26$ &$-22$\\ 
901 & M1\,Iab &2012 Jul 2 & $3795\pm52$ & 0.0 & $0.05\pm0.05$ & 2.16 & $15\:$M$_{\sun}$ & +1 & +6\\
\noalign{\smallskip}
                \hline
        \end{tabular}
	\begin{list}{}{}
\item[]$^{\dagger}$ Spectral type in the spectrum analysed. It appears decidedly later in the lower-resolution spectrum taken in 2007.
\end{list}
\end{table*}

The spectrum of star 70 cannot be properly fit. The Fourier transform indicates two components separated in velocity. Analysis of the spectrum reveals two similar objects separated by about $25\:\mathrm{km}\,\mathrm{s}^{-1}$, i.e. barely resolved at our resolution. This multiplicity could explain the mild spectral variability.
Star 146 cannot be fit either; it is probably too hot for the set of lines used, more suited for K/M spectral types, but fast rotation also helps to dilute the diagnostic lines. We can estimate its rotational velocity even if the resolution of our spectrum is quite low for this task. As we have a good idea of the physical parameters of the star, which must be quite similar to those of star 105, we can choose a suitable stellar model from the POLLUX database \citep{pollux} and convolve it with a gaussian of width appropriate to our spectral resolution. We can then take this synthetic spectrum and convolve it with a rotation profile following \citet{graco09}, and compare the result for different values of $v_\mathrm{rot}$ and limb darkening with our spectrum. This was done following a Bayesian framework, so that we could marginalize over the limb darkening parameter, arriving at a value $v_\mathrm{rot}=53.2\pm1.5\:\mathrm{km}\,\mathrm{s}^{-1}$. This is a very high rotational speed for a supergiant, but we do not see any evidence for a second component in the spectrum.

For all the other stars, we obtain a convincing fit. Due to the low number of lines used in the analysis, the formal uncertainties are moderately large. The values of $T_{\mathrm{eff}}$ found, though, are appropriate for the observed spectral types. Likewise, the analysis of the two spectra of star 126 shows rather better agreement than indicated by the formal uncertainties. Although some of our metallicity determinations have moderate uncertainties, the seven likely members in the cluster core present consistent values. If we average then using the S/N of the spectra as weight, we obtain a mean of $+0.17\pm0.09$~dex, slightly supersolar. This result implies that we can safely use solar-metallicity isochrones for the analysis.

\subsection{Observational HR diagram}
\label{hr}

\begin{figure*}
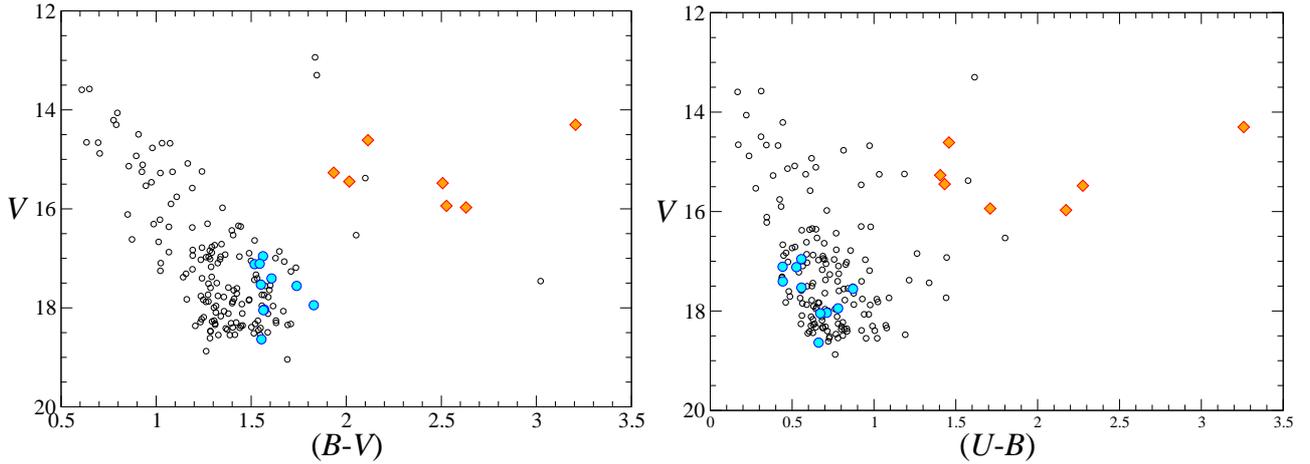

   \centering
 \resizebox{\columnwidth}{!}{\includegraphics[clip]{bv_vs_v.eps}}
 \resizebox{\columnwidth}{!}{\includegraphics[clip]{ub_vs_v.eps}}
      \caption{Raw CMDs for Berkeley~51, from our $UBV$ photometry. The location of the stars observed with GTC and confirmed as early B-type stars is marked by the filled circles. The location of stars observed with the WHT and confirmed as cool supergiants is shown with filled diamonds.
         \label{rawcmds}}
   \end{figure*}

We start the photometric analysis by plotting the $V/(B-V)$ and $V/(U-B)$ diagrams for all stars in the field. In Figure~\ref{rawcmds}, we can observe that the cluster sequence (as defined by the spectroscopically identified early-B stars) is heavily contaminated by what seems to be foreground population. All the confirmed B-type stars have $(B-V)\approx1.5$, and therefore we can assume that this is the location of the cluster sequence. We proceed with a classical photometric analysis, following \citet{jm53} and \citet{johnson58}.

%
%

For early-type stars, it is possible to achieve approximate classification by means of the reddening-free $Q$ parameter, defined as 

 \begin{equation}
\label{Qdefinition}
 Q=(U-B)-\frac{E(U-B)}{E(B-V)}(B-V)
\, ,
\end{equation} 

where $\frac{E(U-B)}{E(B-V)}$ depends on the extinction law \citep[taking a standard value of 0.72;][]{jm53} and also weakly on the spectral type. For a standard extinction law, we can use the expression

\begin{equation}
\label{slope}
\frac{E(U-B)}{E(B-V)}=X+0.05E(B-V)
\, ,
\end{equation} 
where $X$ depends weakly on spectral type (or, correspondingly, intrinsic colour; \citealt{johnson58}), to calculate an accurate $Q$ parameter. Since there are ten early-type members for which we have both classification spectra and $UBV$ photometry (see Table~\ref{intrinsic_members}), we can use them to check if this is a valid approximation. From the spectra, we find that star 131 is a Be star with heavy veiling and anomalous colours, which can thus not be used for this purpose. Star 147 has weak Be characteristics, but still can be left out. For the other eight stars, we calculated $E(U-B)$ and $E(B-V)$ using the intrinsic colours of \citet{fitzgerald}. The average of the reddening slope for them is $0.73\pm0.04$, with the error reflecting the standard deviation. On the other hand, using Eq.~\ref{slope} with the value listed for a typical spectral type B3 in \citet{johnson58}, we find a slope of 0.74. This confirms that the extinction in this direction is standard (see also Section~\ref{sec:ext} later) and that we can use $Q$ with this value of the slope.

Now we use the expression $(B-V)_{0}=0.332\cdot Q$ \citep{johnson58} to estimate the intrinsic colours and hence the colour excess $E(B-V)$ to each star. For the stars with spectra, we can compare the values of $(B-V)_{0}$ obtained from the spectral type calibration with those derived from the $Q$ method. The average difference is $-0.01\pm0.01$~mag, and all the differences are within the typical dispersion of the calibration. The photometric classifications agree with our spectra in placing the main-sequence turnoff at B3.

Application of this method reveals that a large fraction of the stars in our photometry are B-type stars (i.e. $Q<0$). However, many of them have to be foreground to the cluster. For example, star 42 has $Q=-0.61$, corresponding to a mid-B star, but with $V=14.3$ and $(B-V)=0.79$, it is very far away from the position of the spectroscopic members. In addition there is a significant fraction of late-B stars ($0>Q>-0.2$) that have $(B-V)\la1.1$. These objects are in all likelihood foreground interlopers. This high percentage of early-type contaminants is not so surprising as one could na\"{\i}vely think, because the line of sight goes through the Galactic plane, reaching a high distance, and we are selecting only objects with good $U$-band photometry. To confirm this foreground character, we divided the B-type stars in two groups, those with $E(B-V)>1.3$ and those with $E(B-V)\leq1.3$. The former group contains most of the stars with spectral types B3\,--\,B5 and is much more tightly concentrated around the position of the cluster than the latter. So we can start our membership determination by discarding stars with values below this threshold as non-members.

The eight non-emission stars with classification spectra (which had been selected because they seemed to belong to the top of the cluster sequence) have an average $E(B-V)= 1.80$ with a standard deviation $\sigma=0.10$. The standard deviation indicates that the colour excess is only moderately variable for all these likely members. In view of this, we can increase our threshold by rejecting stars with $E(B-V)$ more than 3\,$\sigma$ away from this value as probable non-members. There are no stars with colour excesses more than 2\,$\sigma$ above this average, but many objects with values lower than 3\,$\sigma$ below the average. 

To make the final selection, we proceed with an iterative approach. Firstly, we divide the sample in two groups. Stars with $E(B-V)\geq1.65$ (less than 1.5\,$\sigma$ away from the average for stars with spectra) are taken as likely members, while the rest are taken a possible members. For the first list, we derive photometric spectral types by comparing their $Q$ values to the calibration of \citet{johnson58} for main-sequence stars. We then calculate a first estimate of the extinction by taking the intrinsic colours from \citet{fitzgerald} corresponding to this photometric type, calculating the corresponding colour excess and then applying $A_{V}=3.1\times E(B-V)$. With this value, we deredden all the stars and obtain a dereddened magnitude $m_{V}=V-A_{V}$.

   \begin{figure}
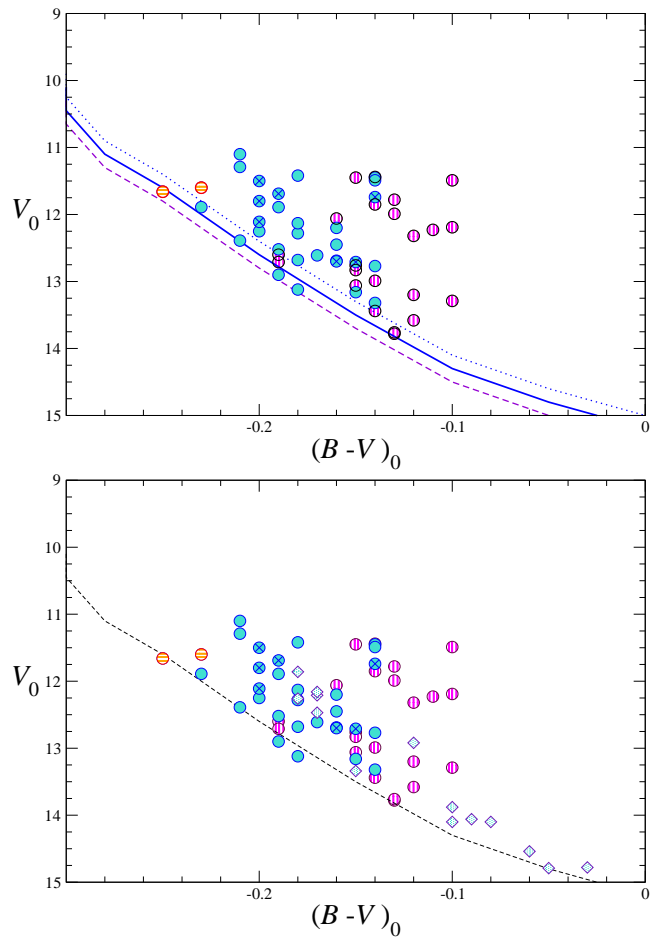

   \centering
\resizebox{\columnwidth}{!}{\includegraphics[angle=0,clip]{zams_final.eps}}
\resizebox{\columnwidth}{!}{\includegraphics[angle=0,clip]{be51_and_Mel20.eps}}
   \caption{\textbf{Top panel:} Dereddened CMD for very likely (filled blue circles) and possible (magente circles with vertical stripes) members of Be~51 with the ZAMS of Schmidt-Kaler displaced at different distance moduli ($\mu=13.5$, 13.7 and 13.9). The two Be stars are marked by orange circles with horizontal stripes. \textbf{Bottom panel: } The same with the confirmed mebers of Melotte~20 displaced to the same $\mu$ overploted as cyan diamonds.\label{zams}}
    \end{figure}

For the second list (stars with $1.47 \leq E(B-V)< 1.65$), we carry out the same procedure separately. We then plot the likely members and the possible members against the observational ZAMS from Schmidt-Kaler \citep{lb6} displaced at different distance moduli (as in Fig.~\ref{zams}). We obtain an initial cluster distance modulus ($\mu$). After subtracting this $\mu$, we compare the absolute magnitude of each star to the photometric spectral type to estimate whether the star has to be a giant or a dwarf to belong to the cluster. Taking into account this luminosity classification, we then proceed to re-estimate the spectral type for the objects that must be giants to be cluster members, using the corresponding $Q$ calibration from \citet{johnson58}. The colour excesses are re-calculated and the process converges, because the differences in colour between giants and main-sequence stars are very small.  We then re-check the fit to the ZAMS and proceed to reject as non-members objects that should have spectral types unexpected for a cluster with a turnoff at B3 (for example, b8\,iv) if they were at the cluster distance. A few objects with spectral type b7 that are too bright to be on the main sequence are moved from the list of likely members to the list of possible members. This does not affect our ZAMS fit, which steadily returns $\mu=13.7$. Figure~\ref{zams} (top panel) shows the fit of the ZAMS to the very likely and possible members that remain after this procedure. The turnoff around $(B-V)_{0}\approx-0.20$ seems to be well defined. The two detected Be stars fall clearly to the left of this turnoff, but this is not unusual for Be stars (and their spetral types show them to be blue stragglers in any case). A third star, \#99, for which we have no spectrum, occupies a similar location, and could also be a Be star.

The fit is only moderately satisfactory, but as illustrated by the other two values of $\mu$ shown, is the best possible. Unfortunately, the position of the ZAMS seems to be determined mainly by stars in the list of possible members. This introduces the doubt of whether our photometry is really reaching deep enough to touch the ZAMS. On the one hand, since stars with types B5\,--\,6\,V are much fainter than the B3\,V and B5\,III objects that we have observed spectroscopically and our photometry is not very deep, it makes sense to assume that we reach only those ZAMS members with rather lower than average reddening (that have thus been included in the list of possible members). On the other hand, the luminosity classes derived from the spectra are consistently lower than those implied from the photometric types at $\mu=13.7$. In fact, if we assume $A_{V}=3.1\times E(B-V)$ and the intrinsic magnitude calibration of \citet{turner80} to derive spectroscopic distances for the blue stars with spectroscopy, we obtain an average $\mu=14.4\pm0.5$, leaving aside again stars 131 and 143. The two values are just compatible within their respective errors (in view of the top panel of Fig.~\ref{zams}, a conservative error of $\pm0.2$~mag is assumed for the visual fit of the ZAMS as a lower envelope, given the uncertainties under discussion), but the difference is quite significant.

To investigate this issue further, in the lower panel of Fig.~\ref{zams}, we have added the members of Melotte~20 (the $\alpha$~Per cluster), a very well studied cluster which also has a MS turn-off at B3\,V. Taking the photometry from \citet{harris56}, we have displaced them from their \textit{Hipparcos} $\mu$ of 6.2~mag \citep{leeuwen09} to $\mu=13.7$ by applying exactly the same dereddening procedure. We notice that the B3\,--\,5\,V stars in Mel~20 (including the MK primary B3\,V standard 29~Per) are well separated from the ZAMS, while in Be~51 we have B3\,V stars all the way down to the ZAMS. This implies that Be~51 is somewhat younger than Mel~20. The confirmed late-B members of Mel~20 trace very well the ZAMS, giving further support to the value that we take as definitive $\mu=13.7\pm0.2$, corresponding to $d=5.5\pm0.5\:$kpc. 

Leaving out the two known Be stars, the average reddening for stars in the list of very likely members is $\left< E(B-V)=1.78\right>$ with a 1-$\sigma$ dispersion of 0.08~mag. This is essentially identical to the values obtained for the spectroscopic members alone, thus confirming the validity of the selection criteria. The objects in this list fulfill all the requisites to be cluster members. Their derived properties are listed in Table~\ref{intrinsic_members}. The list of possible members, given in Table~\ref{intrin_possible}, includes two b3\,v stars that fall together with the very likely members. The membership of these two objects is rather likely, as B3 field stars are rare, and thus suggests that some stars with lower reddening do indeed belong to the cluster population. On the other hand, a number of objects with classifications b5\,--\,7\,iii that are fainter than the B3\,III\,--\,IV stars are most likely interlopers. In any event, we must take into account that errors in $Q$ are $\ga0.05$~mag for the fainter cluster members, and this can imply changes by almost one whole subtype, which could move an object from one list to the other.

\begin{table}
 \centering
 \begin{minipage}{\columnwidth}
  \caption{Intrinsic parameters for all very likely photometric members, ordered by dereddened magnitude. Spectral types are given for the ten stars observed with GTC/OSIRIS and also for \#103.\label{intrinsic_members}}
  \centering
  \begin{tabular}{lccc}
  \hline
\hline
 \noalign{\smallskip}
   Star   & Photometric& Spectral type& $E(B-V)$  \\
  & spectral type$^{,1}$& (if available)& \\
 \noalign{\smallskip}
 \hline
 \noalign{\smallskip}
71	 &b3\,iv   &	    & 1.86\\
249	 &b3\,iv   &	    & 1.84\\
252	 &b3\,iv   &	    & 1.89\\
86	 &b6\,iii  &	    & 1.81\\		
195	 &b6\,iii  &	    & 1.66\\
114	 &b3\,iv   &  B3\,III   & 1.76\\
147	 &b2v	 &  B2\,IVe   & 1.78$^{,2}$\\
131	 &b1v	 &  B2\,Ve    & 1.85$^{,2}$\\
175	 &b3\,iv   &  B3\,IV    & 2.02\\
143	 &b6\,iii  &  B5\,III   & 1.88\\
122	 &b3\,iv   &  B3\,III--IV&1.72 \\
99	 &b2v	 &	    & 1.75\\
103	 &b3\,v	 &  B2\,V$^{,3}$   & 1.79\\
137	 &b3\,v	 &  B3\,IV--V  & 1.75\\
152	 &b3\,v	 &	    & 1.73\\
267	 &b5\,iv   &	    & 1.76\\		
141	 &b3\,v	 &	    & 1.77\\
163	 &b3\,v	 &	    & 1.77\\		
44	 &b3\,v   &	    & 1.91\\
47	 &b5\,v	 &	    & 1.78\\
185	 &b3\,v	 &	    & 1.88\\		
69	 &b5\,v	 &	    & 1.72\\
93	 &b3\,v	 &	    & 1.67\\
177	 &b5\,v	 &	    & 1.75\\
153	 &b5\,v	 &  B3\,V     & 1.73\\
82	 &b5\,v	 &  B5\,III   & 1.72\\
154	 &b6\,iv  &	    & 1.77\\
79	 &b3\,v	 &	    & 1.73\\
116	 &b3\,v	 &	    & 1.72\\
269	 &b5\,v	 &	    & 1.67\\
166	 &b5\,v	&B5\,IV	&1.73\\
169  &b6\,v	& &	1.69	\\
\noalign{\smallskip}
\hline
\end{tabular}
\begin{list}{}{}
\item $^{1}$ Spectral types are derived from the photometry following the procedure described in the text, under the assumption of $\mu=13.7$.
\item $^{2}$ Colour excess values for the confirmed Be star are not entirely interstellar.
\item $^{3}$ Derived from the $Z$-band spectrum, and so less certain.
\end{list}	   
\end{minipage}
\end{table}

\begin{table}
 \centering
 \begin{minipage}{\columnwidth}
  \caption{Intrinsic parameters for stars that could be members, ordered by dereddened magnitude,\label{intrin_possible}}
  \centering
  \begin{tabular}{lcc}
  \hline
\hline
 \noalign{\smallskip}
   Star   &Spectral type$^{,1}$ & $E(B-V)$ \\
 \noalign{\smallskip}
 \hline
 \noalign{\smallskip}
174  & b6\,iii    & 1.59  \\
56   & b5\,iii    & 1.58  \\
30   & b7\,iii    & 1.84  \\
184  & b7\,iii    & 1.53  \\
21   & b6\,iii    & 1.63  \\
9    & b7\,iii    & 1.63  \\
2    & b5\,iii    & 1.50  \\
246  & b7\,iii    & 1.54  \\
31   & b7\,iii    & 1.71  \\
48   &  b7\,iii   &  1.69 \\
239  &  b7\,iv    &  1.77 \\
111  &  b3\,v     &  1.62 \\ 
248  &  b3\,v     &  1.61 \\
136  &  b5\,v     &  1.64 \\ 
102  &  b5\,v     &  1.53 \\
100  &  b6\,v     &  1.59 \\
49   &  b5\,v     &  1.55 \\ 
85   &  b7\,iv    &  1.71 \\
144  &  b7\,iv    &  1.65 \\  
20   &  b5\,v     &  1.57 \\  
14   &  b7\,v     &  1.56 \\  
72   &  b7\,v     &  1.50 \\   
91   &  b7\,v     &  1.50 \\  
\noalign{\smallskip}
\hline
\end{tabular}
\begin{list}{}{}
\item $^{1}$ Spectral types are derived from the photometry following the procedure described in the text. The luminosity class indicated is that needed to be a cluster member, but not necessarily the true value.
\end{list}	   
\end{minipage}
\end{table}
--------------------------------------------------------------------------------------------------

\section{Discussion}

\subsection{Extinction}
\label{sec:ext}

Cluster members with spectroscopy give an average $\left< E(B-V) \right>= 1.80\pm0.10$.
There is a degree of variability in extinction, expected for a distant object in this region of the Galactic Plane. Images of the area (e.g.\ DSS2) reveal the presence of extended nebulosity in this area, and a possible uncatalogued \ion{H}{ii} region about $6\arcmin$ NE from the cluster. WISE images also show substantial dust emission over the whole area. In the optical images, there is a clear contrast between the high stellar density seen to the East and South of the cluster and the much lower density to the West and Northwest, suggestive of a foreground dark cloud. The objects listed in Table~\ref{intrinsic_members} display a moderately broad range of colour excesses, but this is by design (a consequence of the selection procedure in Sect.~\ref{hr}), and so this range should be taken as a lower limit. The list of possible members in Table~\ref{intrin_possible} includes two b3\,v stars (\#111 and \#248) that are very likely to belong to the cluster and present $E(B-V)\approx1.6$, significantly below the range used to define certain members. On the other hand, a number of objects in Table~\ref{intrin_possible} with classifications b5\,--\,7\,iii that are less bright than the stars around the turnoff are unlikely to be members, even though some of them have $E(B-V)>1.7$. This confirms that $E(B-V)$ alone is not enough to identify members, while the range of $E(B-V)$ present in the cluster is likely wider than that adopted to select certain members.

To verify that the reddening law can be approximated by the standard values, we can also calculate the infrared excess for very likely members, using the calibration of intrinsic colours of \citet{winkler97}, the photometric spectral types and $(J-K_{\textrm{S}})$ values from 2MASS. We find $\left< E(J-K_{\textrm{S}}) \right> = 0.89\pm0.06$, in line with the dispersion expected from the optical value (taking into account the relatively large errors of 2MASS for the faintest objects). We find
\begin{equation}
\frac{\left< E(J-K_{\textrm{S}}) \right>}{\left< E(B-V) \right>}=0.50 \, ,
\end{equation}
in good agreement with the expectations for a standard extinction law. As the two averages are obtained with different samples\footnote{Because not all our photometric members have 2MASS photometry passing the quality criteria.}, the average of the ratios $E(J-K_{\textrm{S}})$/$E(B-V)$ for individual targets with good quality photometry, namely $0.49\pm0.03$ (where the error is the standard deviation), is probably more informative, and again ratifies the standard reddening law\footnote{A value of 0.52 is expected for \citet{riekes}.}.

%
   \begin{figure}
   \centering
 \resizebox{\columnwidth}{!}{\includegraphics[clip]{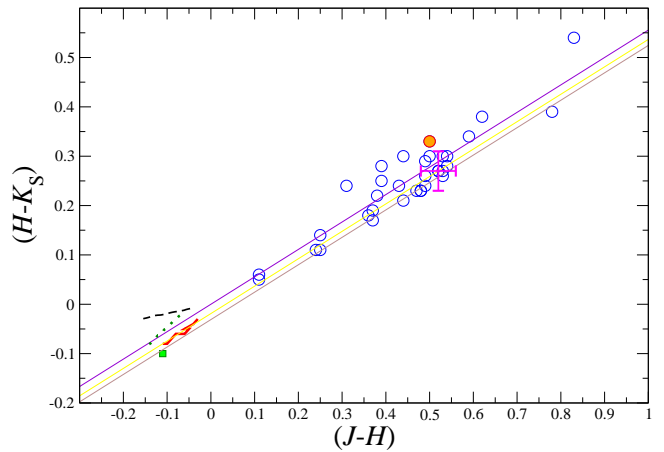}}
       \caption{Colour-colour diagram for early-type stars in the central $3\arcmin$ of the field, selected as in Section~\ref{2mass}. The thin straight lines are reddening vectors for a standard \citet{riekes} extinction law that go through the position of three spectroscopically confirmed cluster members. The thick (red) wavy track is the locus of luminosity-class V B-type stars according to the observational calibration of \citet{winkler97}. The dotted line is the position of stars with masses between $3\:$M$_{\sun}$ and  $12\:$M$_{\sun}$ in a PARSEC \citep{parsec} 3~Ma isochrone (essentially, a ZAMS for B-type stars). The dashed line is the corresponding position for 3\,-- \,$12\:$M$_{\sun}$ in a 3~Ma Geneva \citep{georgy13rot} isochrone. The green square is the colour of all O-type main-sequence stars according to the synthetic models of \citet{marplez06}. The orange circle is the Be star 147. The error bars indicate the typical uncertainty in the photometry.
         \label{extinction}}
   \end{figure}
%

Despite this, an attempt to individually deredden members with good 2MASS photometry using the intrinsic colours of \citet{winkler97} results in positions on the CMD that are too red compared to the isochrone. As this situation has prevented us from individual dereddening of stars in the past \citep[e.g.][]{marco14}, we have carried out an analysis of the infrared reddening law for the field. For this purpose, we take all the stars that had been selected in Section~\ref{2mass} as likely early-type stars with good-quality 2MASS photometry within the central $3\arcmin$ and plot their $(J-H)$ and $(H-K_{\textrm{S}})$ colours, together with the reddening vector, compared to the expected position of early-type stars according to several references in Fig.~\ref{extinction}. As can be seen, the observed position of all the stars lies along the standard reddening vector. It is clear that all stars deproject to positions compatible with the observational colour calibration of \citet{winkler97}, even though they are in the $JHK$ system and some small differences with the 2MASS system are expectable, but not with the theoretical positions predicted by either the Geneva or Padova isochrones, which are in the 2MASS system. We have repeated this experiment with the much younger open cluster Berkeley~90, which shows strong differential reddening \citep{maiz15}, coming to an identical conclusion: in the near-IR colour-colour diagram, early-type stars project along the reddening line to a locus compatible with the empirical colours of B-type star, but not with the colours found in the isochrones. This suggests that the transformations used to convert the isochrones from the theoretical plane to magnitudes and colours do not reproduce well the near-infrared photometry of early-type stars. Note that the issue does not concern the $(J-K_{\textrm{S}})_{0}$ colour, which is quite well reproduced (as can be seen in Fig.~\ref{ircmd}), but the $(J-H)_{0}$ and $(H-K_{\textrm{S}})_{0}$ colours. The first is always too high and the second always too low.

\subsection{Cluster parameters}
\label{params}

\begin{figure}
   \centering
 \resizebox{\columnwidth}{!}{\includegraphics[clip]{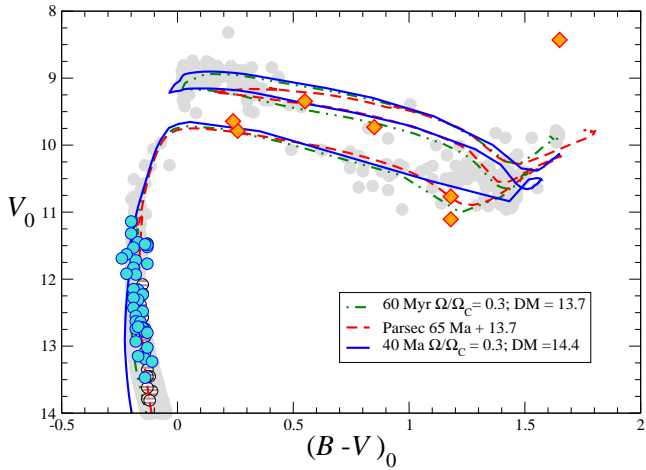}}
       \caption{Dereddened CMD for very likely members (blue filled circles and diamonds) and possible members (brown striped circles) of Be~51. The solid blue line is the $40\:$Ma Geneva \citep{georgy13rot} track displaced to the spectroscopic $\mu=14.4$. The dash-dotted green line is the  $60\:$Ma Geneva track displaced to $\mu=13.7$. The dashed red line is the $65\:$Ma \textsc{parsec} \citep{parsec} isochrone displaced to $\mu=13.7$. The grey points are stars from the synthetic cluster mentioned in the text, generated with the best-fit parameters and the tools of the Geneva group. 
         \label{optcmd}}
   \end{figure}

Using the values from the previous analysis, we plot a dereddened CMD for the cluster in Fig.~\ref{optcmd}. The supergiants have been dereddened following the procedure of \citet{fernie}, which transforms the observed $E(B-V)$ to an equivalent $E(B-V)$ for early type stars to account for colour effects\footnote{This procedure results in $\left< E(B-V) \right> = 1.76\pm0.12$ for seven supergiants with photometry, in perfect agreement with the values for the blue members.}. To determine cluster parameters, we used the isochrones of \citet{georgy13rot} that cover the B-type range at a wide range of initial rotational velocities, $v_{\textrm{rot}}$. The effects of rotation introduce an extra dimension that cannot be constrained with the observed CMD. While the position of the supergiants narrows down significantly the range of possible ages, isochrones with high initial average rotations are almost indistinguishable from younger isochrones with lower initial rotation. For example, isochrones for 30~Ma with no rotation and 40~Ma with a moderately high rotation ($\Omega$/$\Omega_{\textrm{crit}}=0.5$) are almost identical. As a compromise, we will use isochrones with moderate rotation, $\Omega$/$\Omega_{\textrm{crit}}=0.3$. 

%
%

If we accept the value of $\mu=13.7$ that we obtained from the ZAMS fitting, then the isochrone for 60~Ma provides a good fit to the data. If, however, we accept the $\mu=14.4$ that we obtain from the spectral classification, and thus the implication that our photometry might not be deep enough to reach the ZAMS, then the isochrone for 40~Ma provides a similarly good fit. The position of the red supergiants around $(B-V)_{0}=1.1$ may be considered as support for the older value, but we must note that the good agreement with the isochrones must be interpreted with great caution. The evolved stars occupy positions compatible with the isochrones, but not those where the models predict that they should spend most of the time. To illustrate this, we have generated an artificial cluster using the interactive tools made available by the Geneva group\footnote{https://obswww.unige.ch/Recherche/evoldb/index/} \citep{georgy17} and employing the same astrophysical parameters that we use in the fit: $\Omega$/$\Omega_{\textrm{crit}}$=0.3, solar composition and $\tau=60\:$Ma. The cluster initially has 20\,000 intermediate-mass stars, so that statistical sampling is not an issue. This artificial cluster is overplotted on Fig.~\ref{optcmd}, displaying the main properties of the evolutionary tracks: stars spend the first part of the He-burning phase as rather cool red giants (with spectral types K4\,--\,M0) and the second half as blue (A or F) supergiants. None of the evolved stars occupy the regions of highest density in the synthetic cluster. In particular, the model predicts that $\ga$30\% of the He-burning stars should appear as post-RSG A-type supergiants, while we do not see any. 

This discrepancy is not unique to Be~51, but widespread. The similarly-aged cluster Berkeley~55 contains one late-F and five K (super)giants. Except for one, the spectral types of the K super(giants) place them at the same position on the isochrone as the Be~51 objects \citep{negmar12}. There are other well-studied clusters with similar ages, but they do not contain many evolved stars. For example, IC~4665 has no evolved stars, while Melotte~20 only includes the F5\,Ib supergiant $\alpha$~Per. Looking at the literature, we find NGC~4609, NGC~6546 and Trumpler~3 with one red (super)giant each; NGC~6520 and NGC~6649 contain both F-type and red (super)giants, while NGC~6834 and NGC~7654 contain one F supergiant each. Only two clusters of similar age, NGC~5281 and NGC~2345, have known A\,Ib/II stars, but their locations on CMDs are compatible with stars moving off the main sequence, not with a post-RSG nature. Even when we add together all these clusters, the total population is still small. However, the fraction of F-type stars is much higher than predicted by the Geneva models, while post-RSG A supergiants are almost absent. This suggests that the models predict a blue loop that extends to higher temperatures than supported by observations. In fact, if we plot a $65\:$Ma \textsc{parsec} isochrone \citep{parsec} in Fig.~\ref{optcmd}, we see that the only significant difference with the $60\:$Ma moderate-rotation Geneva isochrone is the size of the blue loop, which seems more compatible with the observations. Even then, the two mid-F supergiants in Be~51 occupy positions more consistent with the "Hertzsprung gap" than with the loop.

\begin{figure}
   \centering
 \resizebox{\columnwidth}{!}{\includegraphics[clip]{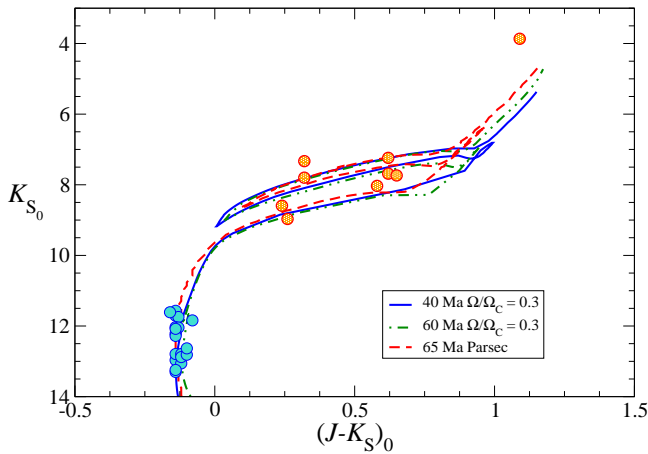}}
       \caption{Dereddened CMD diagram for cluster very likely members with useful 2MASS photometry. The isochrones are as in Fig.~\ref{optcmd}. 
         \label{ircmd}}
   \end{figure}

As a further test, in Fig.~\ref{ircmd} we  plot the individually-dereddened 2MASS data for the supergiants and the few blue members with high-quality photometric values together with the same isochrones used to fit the optical data. For the cool supergiants, we use the average calibration of \citet{gonzalez12} for K and M stars, and the intrinsic colours of F-type supergiants from \citet{koornneef} transformed to the 2MASS system according to the equations of \citet{carpenter01}\footnote{As updated in \newline \small{{\tt http://www.astro.caltech.edu/~jmc/2mass/v3/transformations/}}.}. The match is quite good, but we note that:
\begin{itemize}
 \item The red supergiants lie very far away from their expected position. However, this is likely an artifact of the intrinsic colour calibration, which goes from $(J-K_{\textrm{S}})_{0}=0.58$ at G8\,I to 
$(J-K_{\textrm{S}})_{0}=0.87$ at K1\,I (while it is 1.05 at M1\,I).
\item The position of star 70 is slightly different from its location in the optical diagram. However, this is entirely dependent on the assumption of an F8\,Ib spectral type. The $(B-V)=2.52$ of \#162 is much redder than the $(B-V)=2.11$ of \#70, but both have the same $(J-K_{\textrm{S}})=1.37$. With a spectral type F8\,Ib, \# 70 has an $E(J-K_{\textrm{S}})=1.04$, well above the cluster average. All this suggests that \#70 may have had a later spectral type at the time of the 2MASS observations.
\item The position of star 126 is completely inconsistent with the rest of the cluster in both diagrams.
\end{itemize}
%
%

\begin{figure}
   \centering
 \resizebox{\columnwidth}{!}{\includegraphics[clip]{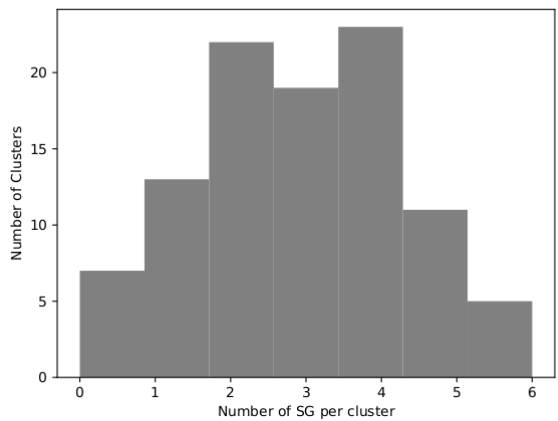}}
       \caption{Number of supergiants found in a sample of 100 synthetic clusters with masses $\sim3\,000\:$M$_{\sun}$. The plot shows the number of clusters containing a given number of supergiants (see Appendix~\ref{synth}). 
         \label{synthsgs}}
   \end{figure}

We cannot give too much weight to the position of any of the individual evolved stars, as post-MS evolution is fast, and stochastic effects must contribute strongly to the very different evolved populations seen in the clusters mentioned above. To quantify this effect, 100 synthetic clusters were generated, all with the same parameters: solar metallicity, an age of $60\:$Ma and a distribution of initial rotational velocities. Each cluster has an initial total mass around $3\,000\:$M$_{\sun}$, i.e.\ typical for a moderately massive Galactic young cluster. Details of the clusters are presented in Appendix~\ref{synth}. The number of supergiants in a given cluster ranges from zero (7 out of 100 clusters) to six (5 out of 100 clusters). The distribution is shown in Fig.~\ref{synthsgs}. Most clusters have between two and four evolved stars (about 20\% in each case) and the average number is three supergiants per cluster. This suggests that Be~51 must have a higher initial mass, at least twice as much to be in the range of statistical probability and most likely three times more, i.e.\ between $6$ and $9\times10^{3}\:$M$_{\sun}$.

%
%

\subsection{Radial velocities}

Radial velocities were calculated following the procedure outlined in \citet{kopo11}. Since ISIS is attached to the Cassegrain focus of the WHT, it moves with the telescope, and is subject to large flexures. The method employed uses sky emission to refine possible systematics remaining in the wavelength calibration, so that all the spectra are in a common system that can then be anchored by using velocity standards. After this, it compares the observed spectra with a battery of models using a Bayesian framework. This has the advantage that it is possible to marginalize over any parameter in which we are not interested, removing it from the analysis while at the same time taking it into account when deriving uncertainties. In our case, we marginalize over the continuum normalization, since continuum determination is almost impossible at these resolutions for late type stars, and over the stellar model, so that the derived velocities are not model dependent. According to the ISIS manual, an internal accuracy of $2\:$km\,s$^{-1}$ can be achieved with R1200R, though a small systematic shift could be present. Two observations of star 126 taken on two separate nights (4 nights apart) differ by $3\:$km\,s$^{-1}$, suggesting that the manual does not overestimate greatly the accuracy achievable. The accuracy is likely to be lower for the F-type stars, which have less (and generally broader) lines in the range used. The radial velocities measured are listed in Table~\ref{partab}.

At first sight, the radial velocities do not seem consistent with a single population. The average of all the values (including the two measurements of \#126) is $v_{\textrm{LSR}}=-8\:$km\,s$^{-1}$, with a standard deviation of $5\:$km\,s$^{-1}$. However, we can see that there are only three stars with velocities moderately different from the rest. Of these, two are the objects identified as spectroscopic variable, stars 70 and 162. If these objects are on the instability strip or binary (as seems to be the case of \#70), radial velocity variations are expected. If we leave out these two stars, there is only one outlier, \#172. This object could be in a binary or perhaps have recently been ejected from the cluster. If we also ignore it, the other six objects give an average of $v_{\textrm{LSR}}=-4\:$km\,s$^{-1}$, with a standard deviation of $2\:$km\,s$^{-1}$, perfectly compatible with the expected internal accuracy of the measurements. 

We calculated the Galactic rotation curve with respect to the local standard of rest (LSR) in the direction of Be~51 using the fit of \citet{reid14}, with $R_{0}=8.34\:$kpc and $\phi_{0}=252.2\:$km\,s$^{-1}$. A cluster velocity of $v_{\textrm{LSR}}=-4\:$km\,s$^{-1}$ corresponds to a kinetic distance of $\approx 5.5\:$kpc, in very good agreement with the photometric distance. A longer distance of $7.5\:$kpc would correspond to a radial velocity $v_{\textrm{LSR}}\approx-25\:$km\,s$^{-1}$. Therefore the observed radial velocity favours the distance derived from the photometric solution. At a distance of $\sim5.5$~kpc and with $\ell \approx72\degr$, Be~51 would be placed behind the whole extent of the Local arm, which explains its high and patchy extinction. \citet{xu13} show that the Local arm is a major structure, consisting of several important star-forming regions and extending until at least 4~kpc away from the Sun in this direction. However, at this distance Be~51 would already be in the Perseus arm. Indeed, for a distance $\ga6.5$~kpc, Galactic structure models would place the cluster beyond the Perseus arm, in the inter-arm region. This again favours the distance derived from photometry and the older age.

\citet{xu13} assume that the Local arm bends slightly, and branches from the Perseus arm at $\ell\sim55\degr$, but we note that a shallower arm curvature would mean a branching point not far from the position of Be~51. There are no known tracers of the Perseus arm between $\ell\approx50\degr$ and $\ell\approx95\degr$ \citep{choi14}, and so Be~51 could represent an important anchoring point for this arm.

\subsection{The extent of Berkeley~51}

Even if our photometry is deep enough to sample the ZAMS, we are only confidently reaching a spectral type B5\,V, corresponding to a $M_{V}\approx-0.8$, and still are unlikely to be complete for the faintest (i.e.\ more reddened) objects. Even so, we find close to 40 photometric members and a few more possible members down to slightly fainter $M_{V}$. If we assume the older age of $60\:$Ma, these objects are in the range between $\sim4.5\:$M$_{\sun}$ and $\sim6.8\:$M$_{\sun}$, i.e.\ they contribute $\ga200\:$M$_{\sun}$ to the cluster mass. We cannot attempt to derive an initial mass function or a total mass, but by integrating a simple Salpeter law under the assumption that there are 40 stars between $4.5\:$M$_{\sun}$ and $6.8\:$M$_{\sun}$, we find a current mass of $\sim1\,300\:$M$_{\sun}$ in stars more massive than the Sun only, simply from this lower limit. This would imply an initial cluster mass $\ga2\,000\:$M$_{\sun}$.  Integration of a standard \citep{kroupa01} IMF would result in a cluster mass $\approx4\,000\:$M$_{\sun}$. This is a lower limit that does not take into account the incompleteness of our photometry, the effects of binarity or dynamical ejections from the cluster. It is clear that Be~51 is at least a moderately massive cluster. As discussed in Sect.~\ref{params}, comparison to synthetic clusters suggest that a total mass above $5\times10^{3}\:$M$_{\sun}$ is needed for the presence of nine evolved stars.

Even though the cluster appears strongly concentrated, there are some very likely members over the whole field covered by our photometry. In addition, even within small field, there is an indication of increasing reddening towards the North. As mentioned in Sect.~\ref{sec:ext}, the wide field DSS and 2MASS images strongly confirm this impression. 
As a further check on the extent of the cluster, we took spectra of some of the luminous red star candidates found outside the cluster core (see Fig.~\ref{fig:raw} and Table~\ref{log}). Since we could not observe all objects in the diagram, we selected two stars that lie close ($r<3\arcmin$) to the cluster but have colours different from those of cluster supergiants, and two stars found at larger distances, with colours similar to cluster members.

The two nearby stars, 501 (2MASS J20114667+3423097) and 502 (2MASS J20114564+3422422), have spectral types M4\,Ib and M4\,II, respectively. Since they were observed only in the 2007 run, we have no measurement of their radial velocities. The position of star 502 in the CMD clearly rules out an association with the cluster, as it is fainter in $K_{\textrm{S}}$ than the early-K supergiants. Star 501 also appears too faint, and has a colour excess $E(J-K_{\textrm{S}})=0.67$, lower than cluster members. We thus conclude that these two objects are chance projections. 

Star 503 (2MASS J20121264+3420366) lies about $5\arcmin$ southeast of the cluster. It has exactly the same colours and magnitudes as the clump of early K supergiants, and its spectral type K2\,Ib is typical of this group. Its observed radial velocity, however, $v_{\mathrm{ LSR}}=-22\:\textrm{km}\,\textrm{s}^{-1}$, is quite different from the cluster average, even though its metallicity is compatible with the cluster average. We note that there is a moderate number of objects with similar colours and magnitudes within $7\arcmin$ of the cluster that would be worth checking for radial velocities.

Finally, star 901 (2MASS 20115472+3427464) is a very bright infrared source (IRAS 20099+3418) located $\sim3\farcm5$ north of the cluster. It is an M1\,Iab supergiant, with a very high colour excess, $E(J-K_{\textrm{S}})=1.37$, very likely due to the dark cloud discussed above. Its WISE colours have a very large uncertainty, e.g.\ (W1$-$W3)$= 0.13\pm0.26$, but do not suggest a strong intrinsic reddening, associated with heavy mass loss. Its nature will be discussed in Sect.~\ref{sec:msgs}.

\subsection{Comparison to previous work}

We find a younger age than the few previous works on Berkeley~51. Even though our age estimate is fully supported by the spectral types of the stars observed, it is interesting to consider the reasons why Berkeley~51 had not been identified as a young open cluster before. In the case of \citet{tadross}, he obtains an age of 150~Ma, based on 2MASS data. Since the global shape of the isochrones at these ages is similar, the difference is likely due to the distance/age/extinction degeneracy. However, we note that the ages of all young open clusters seem to be overestimated in \citet{tadross}. For example, he correctly identifies Berkeley~90 as the youngest cluster in his sample, but he assigns an age of $\sim100~$Ma, while it is in fact a very young open cluster containing early-O stars \citep{marco17}. This is probably because the only young cluster in his calibration set, Berkeley~55, is given a rather old age of 300~Ma \citep[cf.][]{negmar12}.

A direct comparison of our photometry with that of \citet{subram10} is difficult, as they do not provide coordinates for their objects. However, we can compare $\sim20$ bright objects that are easily identifiable in the cluster chart available in the WEBDA database. There are significant systematic differences between the two photometric datasets. Their $V$ magnitudes are between 0.4 and 0.6~mag brighter than ours, while their $(B-V)$ colours are larger, with their $B$ magnitudes only 0.1 or 0.2~mag brighter than ours. Despite this large difference, inspection of their fig.~22 suggests that the much older age that they give to the cluster is mainly due to an incorrect identification of cluster members, as none of the bright supergiants falls on the isochrone chosen.

\subsection{The yellow supergiants}
Berkeley~51 is remarkable in containing four F-type supergiants. Only the starburst cluster Westerlund~1 has a larger population (6 yellow supergiants), but these are much more massive stars, with $\sim40\:$M$_{\sun}$ \citep{clark10}. NGC~7790, which is somewhat older than Be~51 \citep[$\sim100\:$Ma;][]{majaess13}, hosts three, all of which are Cepheid variables, the two components of the binary CE~Cas, and CF~Cas. NGC~129, with an age similar to NGC~7790, contains the Cepheid DL~Cas and the non-variable F5\,Ib supergiant HD~236433. The Cepheid V376~Cas could be a halo member as well \citep{anderson13}.

We can use the strength of the \ion{O}{i}~7774\AA\ triplet in the spectra of the F-type supergiants to obtain an independent estimation of the cluster distance, by utilising the calibration of \citet{arellano03}. We measured the strength of the triplet (EW74) on the low-resolution 2007 spectra. The triplet is not resolved and we estimate an uncertainty of $\pm0.1$\AA, owing to the definition of the continuum. For star 105, we measure an EW$=1.0\,$\AA, which would correspond to $M_{V}=-4.9$. For \#146, we measure $1.1\,$\AA, i.e. the same value within errors, in good agreement with the fact that the two stars have approximately the same magnitudes and colours in all photometric diagrams. This luminosity is a bit higher than those of $\alpha$~Per or HD~236433, which are quite similar to each other according to \citet{kovt12}, and have about the same spectral type as the Be~51 objects. 

For star 162, however, we measure an EW=$0.6\,$\AA. According to the calibration, this implies an $M_{V}$ of only $-3.0$. For \#70, we measure a similar EW=$0.7\,$\AA, but this object is likely a binary. Given the position of the two late-F stars in the CMDs, we would expect them to be slightly brighter than the mid-F supergiants, instead of fainter. The most likely explanation resides in the fact that the calibration of \citet{arellano03} does not take into account the dependence of EW74 with effective temperature. According to \citet{kovt12}, supergiants of later types have weaker EWs at a given luminosity\footnote{Note that we have not used the calibration of \citet{kovt12} because it has a strong dependence on $\log\,g$ and especially $\xi$, which we have had to fix at assumed values.}. Moreover, as seen in fig.~3 of \citet{kovt12}, for temperatures $\sim 6000\:$K and lower, i.e. as we move into G-types, EW74 seems to show a weaker dependence on luminosity. Star 162 had a spectral type G1\,Ib in the spectrum on which EW74 was measured, and therefore falls in this regime. So we have reason to believe that its $M_{V}$ has been underestimated. We can thus conclude that the values found are consistent, within the uncertainties typical of the calibration (around 0.7~mag), with the isochrones. For the short distance that we have preferred, \#105, \#146 and \#162 have $M_{V}\approx-4.0$, with \#70 somewhat brighter, in agreement with its suspect binary nature.  In fact, we would expect that the four yellow supergiants in Be~51, together with those in other clusters containing more than one such object (NGC~129 and NGC~7790 mentioned above), can help improve the calibration, especially once precise \textit{Gaia} distances exist for all of them.

\subsection{The nature of IRAS 20100+3415}
\label{sec:msgs}

Star 126 is by far the brightest cluster member in the near infrared. It is also the counterpart of the mid-infrared source IRAS 20100+3415. This object, located in the inner core of the cluster, has a radial velocity fully compatible with other cluster members. However, its late spectral type and position in the CMDs is incompatible with the best-fit isochrones. As an M2\,Iab supergiant, it is expected to be a moderately massive star (typically, of $\sim15\:$M$_{\sun}$). Such objects are extremely rare and a chance coincidence with a young open cluster is very unlikely. Moreover, its luminosity is fully consistent with the expectations for an object of this spectral type at the cluster distance, while the observed $v_{\textrm{rad}}$ makes chance coincidence even more unlikely.

Its WISE colours are poorly constrained, probably due to saturation, but with (W1$-$W3$)=0.85\pm0.29$, there is clear evidence of strong mass loss. The interpretation of this object is further compounded by the detection of a second luminous supergiant, star 901 discussed above, only $\sim3\farcm5$ to the North. While K-type supergiants in the Milky Way are generally of luminosity class Ib and can be descended from stars of only 7 or $8\:$M$_{\sun}$ \citep[c.f.][]{negmar12,alonso17}, M-type supergiants of luminosity class Iab are high-mass stars. Given their short lifetimes, they are rare objects (the known Galactic population runs into the several hundred, with estimates of a total population of a few thousand). Except in regions of intense recent star formation, the chance detection of two such objects within $3\farcm5$ is very unlikely \citep[see][for estimates based on actual observations]{neg16multi}. If we take into account that \#126 and \#901 have almost identical \textit{dereddened} colours and magnitudes, the possibility that they are not physically related, in spite of a difference in radial velocity of $13\:$km\,s$^{-1}$ seems very unlikely.

If the cluster age lies in the young half of the range considered, the most evolved stars are expected to have, according to isochrones, $\ga8\:$M$_{\sun}$. We should then consider the possibility that an object like \#126 is a super-AGB star. The exact mass range at which these objects occur depends on the internal physics \citep{poelarends08} and their expected observational properties are unconstrained. To explore this option, we obtained a high-resolution spectrum of \#126 with HERMES, and we scanned its spectrum in search of any indication of an AGB nature, such as the presence of the \ion{Li}{i}~6708\,\AA\ doublet or the \ion{Rb}{i}~7800\,\AA\ line \citep[e.g.][]{anibal07}, without finding any of them. We note that the sample of massive AGB stars where these features have been detected have much later spectral types than M2. The spectrum of \#126, though, seems indistinguishable from those of other \textit{bona fide} red supergiants of the same spectral type that we had observed at the same resolution. In view of this, we consider that the most likely explanation for the presence of \#126 in the cluster is that it is really a more massive star that has formed via mass transfer in an interacting binary. Star 901 would require a more complex explanation, though. Accurate distances to the brightest cluster members in the \textit{Gaia} final release will be able to ascertain this hypothesis.

\section{Conclusions}
We have carried out a comprehensive spectroscopic and photometric study of the highly reddened open cluster Berkeley~51. Our analysis conclusively shows that this is a young open cluster with an important population of evolved stars. We identify a main-sequence turn-off at spectral type B3\,V and at least two Be stars with earlier spectral types. In addition, we find four yellow and five red supergiants in the central overdensity. Two of the yellow supergiants show spectral variability, displaying spectral types F8\,Ib and later, a behaviour typical of Cepheid variables.

A fit to the main sequence indicates a distance of 5.5~kpc, although we may be missing some of the faintest (most heavily reddened) members. The spectral types of some of the brightest members may favour a higher distance, but both the cluster average radial velocity (when compared to the Galactic rotation curve in this direction) and the strength of the \ion{O}{i} triplet in the four yellow supergiants identified support a distance not much higher than 5.5~kpc, which is compatible with a location in the Perseus arm according to most models, even though no other tracers are known in this direction.

Isochrone fits would suggest an age of $\sim 60\:$Ma for moderate-rotation Geneva models or $65\:$Ma for \textsc{parsec} models. Although the isochrones reproduce well the overall distribution of evolved  of stars in the CMDs, the supergiants are not located at positions where the models predict that stars should spend a substantial amount of time. The red supergiants (with spectral types G8 or K0) appear all somewhat warmer than the predictions for the first part of He core burning. Geneva models predict that, after this phase, stars will move to high temperatures and will spend the rest of He core burning as A-type supergiants. The populations observed in a number of cluster with ages $\sim50\:$Ma do not seem to agree with this prediction, as they contain preferentially F-type supergiants, either Cepheid variables or objects with stable spectral type close to F5\,Ib. The \textsc{parsec} isochrones predict shorter blue loops at a given metallicity, which perhaps are in better agreement with observations.

The mid-infrared source IRAS 20100+3415, located near the centre of the cluster, is an M2\,Iab supergiant, probably the descendant of a blue straggler formed via binary interaction. Its WISE colours suggest heavy mass loss. Searches for associated maser emission would be of high interest to exploit the availability of Be~51 as a tracer of the Perseus arm in a poorly known region of the Milky Way.

\section*{Acknowledgements}
We are very thankful to Dr. Sylvia Ekstr\"{o}m for generating the synthetic clusters used in the analysis. We also thank the anonymous referee for constructive comments that improved the paper.

 During part of this work, IND was a visitor at the Institute of Astronomy, University of Cambridge, whose warm hospitality is heartily acknowledged. This visit was funded by the Conselleria de Educaci\'on, Cultura y Deporte of the Generalitat Valenciana under grant BEST/2014/276. This research is partially supported by the Spanish Government Ministerio de Econom\'{\i}a y Competitivad (MINECO/FEDER) under grant AYA2015-68012-C2-2-P. HMT acknowledges support from MINECO under fellowship FJCI-2014-23001. MM acknowledges the support of a research grant funded by the STFC (ST/M001008/1).
 
The photometric observations were obtained with the the Nordic Optical Telescope, operated by the Nordic Optical Telescope Scientific Association. The spectroscopic observations were obtained with the WHT, which is operated on the island of La Palma by the Isaac Newton Group, the Gran Telescopio Canarias (GTC), and the Mercator Telescope, operated by the Flemish Community. All these telescopes are installed in the Spanish Observatorio del Roque de Los Muchachos of the Instituto de Astrof\'{\i}sica de Canarias. The Starlink software (Currie et al.\ 2014) is currently supported by the East Asian Observatory.

 This research has made use of the Simbad, Vizier and Aladin services developed at the Centre de Donn\'ees Astronomiques de Strasbourg, France.  This research has made use of the WEBDA database, operated at the Department of Theoretical Physics and Astrophysics of the Masaryk University. It also makes use of data products from 
the Two Micron All Sky Survey, which is a joint project of the University of
Massachusetts and the Infrared Processing and Analysis
Center/California Institute of Technology, funded by the National
Aeronautics and Space Administration and the National Science
Foundation. This paper makes use of data obtained from the Isaac Newton Group Archive which is maintained as part of the CASU Astronomical Data Centre at the Institute of Astronomy, Cambridge. 




\bibliographystyle{mnras}
\bibliography{clusters} 

\begin{thebibliography}{}
\makeatletter
\relax
\def\mn@urlcharsother{\let\do\@makeother \do\$\do\&\do\#\do\^\do\_\do\%\do\~}
\def\mn@doi{\begingroup\mn@urlcharsother \@ifnextchar [ {\mn@doi@}
  {\mn@doi@[]}}
\def\mn@doi@[#1]#2{\def\@tempa{#1}\ifx\@tempa\@empty \href
  {http://dx.doi.org/#2} {doi:#2}\else \href {http://dx.doi.org/#2} {#1}\fi
  \endgroup}
\def\mn@eprint#1#2{\mn@eprint@#1:#2::\@nil}
\def\mn@eprint@arXiv#1{\href {http://arxiv.org/abs/#1} {{\tt arXiv:#1}}}
\def\mn@eprint@dblp#1{\href {http://dblp.uni-trier.de/rec/bibtex/#1.xml}
  {dblp:#1}}
\def\mn@eprint@#1:#2:#3:#4\@nil{\def\@tempa {#1}\def\@tempb {#2}\def\@tempc
  {#3}\ifx \@tempc \@empty \let \@tempc \@tempb \let \@tempb \@tempa \fi \ifx
  \@tempb \@empty \def\@tempb {arXiv}\fi \@ifundefined
  {mn@eprint@\@tempb}{\@tempb:\@tempc}{\expandafter \expandafter \csname
  mn@eprint@\@tempb\endcsname \expandafter{\@tempc}}}

\bibitem[\protect\citeauthoryear{{Aller} et~al.,}{{Aller} et~al.}{1982}]{lb6}
{Aller} L.~H.,  et~al., eds, 1982, {Landolt-B{\"o}rnstein: Numerical Data and
  Functional Relationships in Science and Technology - New Series ``
  Gruppe/Group 6 Astronomy and Astrophysics '' Volume 2 Schaifers/Voigt:
  Astronomy and Astrophysics / Astronomie und Astrophysik `` Stars and Star
  Clusters / Sterne und Sternhaufen}

\bibitem[\protect\citeauthoryear{{Alonso-Santiago}, {Negueruela}, {Marco},
  {Tabernero}, {Gonz{\'a}lez-Fern{\'a}ndez}  \& {Castro}}{{Alonso-Santiago}
  et~al.}{2017}]{alonso17}
{Alonso-Santiago} J.,  {Negueruela} I.,  {Marco} A.,  {Tabernero} H.~M.,
  {Gonz{\'a}lez-Fern{\'a}ndez} C.,   {Castro} N.,  2017, \mn@doi [\mnras]
  {10.1093/mnras/stx783}, \href
  {http://adsabs.harvard.edu/abs/2017MNRAS.469.1330A} {469, 1330}

\bibitem[\protect\citeauthoryear{{Anderson}, {Eyer}  \& {Mowlavi}}{{Anderson}
  et~al.}{2013}]{anderson13}
{Anderson} R.~I.,  {Eyer} L.,   {Mowlavi} N.,  2013, \mn@doi [\mnras]
  {10.1093/mnras/stt1160}, \href
  {http://adsabs.harvard.edu/abs/2013MNRAS.434.2238A} {434, 2238}

\bibitem[\protect\citeauthoryear{{Arellano Ferro}, {Giridhar}  \& {Rojo
  Arellano}}{{Arellano Ferro} et~al.}{2003}]{arellano03}
{Arellano Ferro} A.,  {Giridhar} S.,   {Rojo Arellano} E.,  2003, \rmxaa, \href
  {http://adsabs.harvard.edu/abs/2003RMxAA..39....3A} {39, 3}

\bibitem[\protect\citeauthoryear{{Barklem}, {Piskunov}  \& {O'Mara}}{{Barklem}
  et~al.}{2000}]{bar00}
{Barklem} P.~S.,  {Piskunov} N.,   {O'Mara} B.~J.,  2000, \mn@doi [\aaps]
  {10.1051/aas:2000167}, \href
  {http://adsabs.harvard.edu/abs/2000A%26AS..142..467B} {142, 467}

\bibitem[\protect\citeauthoryear{{Bressan}, {Marigo}, {Girardi}, {Salasnich},
  {Dal Cero}, {Rubele}  \& {Nanni}}{{Bressan} et~al.}{2012}]{parsec}
{Bressan} A.,  {Marigo} P.,  {Girardi} L.,  {Salasnich} B.,  {Dal Cero} C.,
  {Rubele} S.,   {Nanni} A.,  2012, \mn@doi [\mnras]
  {10.1111/j.1365-2966.2012.21948.x}, \href
  {http://adsabs.harvard.edu/abs/2012MNRAS.427..127B} {427, 127}

\bibitem[\protect\citeauthoryear{{Carpenter}}{{Carpenter}}{2001}]{carpenter01}
{Carpenter} J.~M.,  2001, \mn@doi [\aj] {10.1086/320383}, \href
  {http://adsabs.harvard.edu/abs/2001AJ....121.2851C} {121, 2851}

\bibitem[\protect\citeauthoryear{{Cenarro}, {Cardiel}, {Gorgas}, {Peletier},
  {Vazdekis}  \& {Prada}}{{Cenarro} et~al.}{2001}]{cenarro01}
{Cenarro} A.~J.,  {Cardiel} N.,  {Gorgas} J.,  {Peletier} R.~F.,  {Vazdekis}
  A.,   {Prada} F.,  2001, \mn@doi [\mnras] {10.1046/j.1365-8711.2001.04688.x},
  \href {http://adsabs.harvard.edu/abs/2001MNRAS.326..959C} {326, 959}

\bibitem[\protect\citeauthoryear{{Chiosi}, {Bertelli}  \& {Bressan}}{{Chiosi}
  et~al.}{1992}]{chiosi}
{Chiosi} C.,  {Bertelli} G.,   {Bressan} A.,  1992, \mn@doi [\araa]
  {10.1146/annurev.aa.30.090192.001315}, \href
  {http://adsabs.harvard.edu/abs/1992ARA%26A..30..235C} {30, 235}

\bibitem[\protect\citeauthoryear{{Choi}, {Hachisuka}, {Reid}, {Xu},
  {Brunthaler}, {Menten}  \& {Dame}}{{Choi} et~al.}{2014}]{choi14}
{Choi} Y.~K.,  {Hachisuka} K.,  {Reid} M.~J.,  {Xu} Y.,  {Brunthaler} A.,
  {Menten} K.~M.,   {Dame} T.~M.,  2014, \mn@doi [\apj]
  {10.1088/0004-637X/790/2/99}, \href
  {http://adsabs.harvard.edu/abs/2014ApJ...790...99C} {790, 99}

\bibitem[\protect\citeauthoryear{{Clark}, {Ritchie}  \& {Negueruela}}{{Clark}
  et~al.}{2010}]{clark10}
{Clark} J.~S.,  {Ritchie} B.~W.,   {Negueruela} I.,  2010, \mn@doi [\aap]
  {10.1051/0004-6361/200913820}, \href
  {http://adsabs.harvard.edu/abs/2010A%26A...514A..87C} {514, A87}

\bibitem[\protect\citeauthoryear{{Comer{\'o}n} \& {Pasquali}}{{Comer{\'o}n} \&
  {Pasquali}}{2005}]{cp05}
{Comer{\'o}n} F.,  {Pasquali} A.,  2005, \mn@doi [\aap]
  {10.1051/0004-6361:20041788}, \href
  {http://adsabs.harvard.edu/abs/2005A%26A...430..541C} {430, 541}

\bibitem[\protect\citeauthoryear{{Currie}, {Berry}, {Jenness}, {Gibb}, {Bell}
  \& {Draper}}{{Currie} et~al.}{2014}]{currie14}
{Currie} M.~J.,  {Berry} D.~S.,  {Jenness} T.,  {Gibb} A.~G.,  {Bell} G.~S.,
  {Draper} P.~W.,  2014, in {Manset} N.,  {Forshay} P.,  eds,  Astronomical
  Society of the Pacific Conference Series Vol. 485, Astronomical Data Analysis
  Software and Systems XXIII. p.~391

\bibitem[\protect\citeauthoryear{{Diaz}, {Terlevich}  \& {Terlevich}}{{Diaz}
  et~al.}{1989}]{diaz89}
{Diaz} A.~I.,  {Terlevich} E.,   {Terlevich} R.,  1989, \mnras, \href
  {http://adsabs.harvard.edu/abs/1989MNRAS.239..325D} {239, 325}

\bibitem[\protect\citeauthoryear{{Dorda}, {Gonz{\'a}lez-Fern{\'a}ndez}  \&
  {Negueruela}}{{Dorda} et~al.}{2016}]{dor16b}
{Dorda} R.,  {Gonz{\'a}lez-Fern{\'a}ndez} C.,   {Negueruela} I.,  2016, \mn@doi
  [\aap] {10.1051/0004-6361/201628422}, \href
  {http://adsabs.harvard.edu/abs/2016A%26A...595A.105D} {595, A105}

\bibitem[\protect\citeauthoryear{{Draper}, {Taylor}  \& {Allan}}{{Draper}
  et~al.}{2011}]{draper}
{Draper} P.~W.,  {Taylor} M.,   {Allan} A.,  2011, Starlink User Note, \href
  {http://adsabs.harvard.edu/abs/2011StaUN.139.....D} {139}

\bibitem[\protect\citeauthoryear{{Dutra-Ferreira}, {Pasquini}, {Smiljanic},
  {Porto de Mello}  \& {Steffen}}{{Dutra-Ferreira} et~al.}{2016}]{dut16}
{Dutra-Ferreira} L.,  {Pasquini} L.,  {Smiljanic} R.,  {Porto de Mello} G.~F.,
   {Steffen} M.,  2016, \mn@doi [\aap] {10.1051/0004-6361/201526783}, \href
  {http://adsabs.harvard.edu/abs/2016A%26A...585A..75D} {585, A75}

\bibitem[\protect\citeauthoryear{{Ekstr{\"o}m} et~al.,}{{Ekstr{\"o}m}
  et~al.}{2012}]{ekstrom12}
{Ekstr{\"o}m} S.,  et~al., 2012, \mn@doi [\aap] {10.1051/0004-6361/201117751},
  \href {http://adsabs.harvard.edu/abs/2012A%26A...537A.146E} {537, A146}

\bibitem[\protect\citeauthoryear{{Ekstr{\"o}m}, {Georgy}, {Meynet}, {Groh}  \&
  {Granada}}{{Ekstr{\"o}m} et~al.}{2013}]{ekstrom13}
{Ekstr{\"o}m} S.,  {Georgy} C.,  {Meynet} G.,  {Groh} J.,   {Granada} A.,
  2013, in {Kervella} P.,  {Le Bertre} T.,   {Perrin} G.,  eds,  EAS
  Publications Series Vol. 60, EAS Publications Series. pp 31--41 (\mn@eprint
  {arXiv} {1303.1629}), \mn@doi{10.1051/eas/1360003}

\bibitem[\protect\citeauthoryear{{Fernie}}{{Fernie}}{1963}]{fernie}
{Fernie} J.~D.,  1963, \mn@doi [\aj] {10.1086/109215}, \href
  {http://adsabs.harvard.edu/abs/1963AJ.....68..780F} {68, 780}

\bibitem[\protect\citeauthoryear{{Fitzgerald}}{{Fitzgerald}}{1970}]{fitzgerald}
{Fitzgerald} M.~P.,  1970, \aap, \href
  {http://adsabs.harvard.edu/abs/1970A%26A.....4..234F} {4, 234}

\bibitem[\protect\citeauthoryear{{Foreman-Mackey}, {Hogg}, {Lang}  \&
  {Goodman}}{{Foreman-Mackey} et~al.}{2013}]{fma13}
{Foreman-Mackey} D.,  {Hogg} D.~W.,  {Lang} D.,   {Goodman} J.,  2013, \mn@doi
  [\pasp] {10.1086/670067}, \href
  {http://adsabs.harvard.edu/abs/2013PASP..125..306F} {125, 306}

\bibitem[\protect\citeauthoryear{{Garc{\'{\i}}a-Hern{\'a}ndez},
  {Garc{\'{\i}}a-Lario}, {Plez}, {Manchado}, {D'Antona}, {Lub}  \&
  {Habing}}{{Garc{\'{\i}}a-Hern{\'a}ndez} et~al.}{2007}]{anibal07}
{Garc{\'{\i}}a-Hern{\'a}ndez} D.~A.,  {Garc{\'{\i}}a-Lario} P.,  {Plez} B.,
  {Manchado} A.,  {D'Antona} F.,  {Lub} J.,   {Habing} H.,  2007, \mn@doi
  [\aap] {10.1051/0004-6361:20065785}, \href
  {http://adsabs.harvard.edu/abs/2007A%26A...462..711G} {462, 711}

\bibitem[\protect\citeauthoryear{{Georgy} \& {Ekstr{\"o}m}}{{Georgy} \&
  {Ekstr{\"o}m}}{2017}]{georgy17}
{Georgy} C.,  {Ekstr{\"o}m} S.,  2017, in {Charbonnel} C.,  {Nota} A.,  eds,
  IAU Symposium Vol. 316, Formation, Evolution, and Survival of Massive Star
  Clusters. pp 355--356 (\mn@eprint {arXiv} {1509.02779}),
  \mn@doi{10.1017/S1743921315008868}

\bibitem[\protect\citeauthoryear{{Georgy}, {Ekstr{\"o}m}, {Granada}, {Meynet},
  {Mowlavi}, {Eggenberger}  \& {Maeder}}{{Georgy} et~al.}{2013}]{georgy13rot}
{Georgy} C.,  {Ekstr{\"o}m} S.,  {Granada} A.,  {Meynet} G.,  {Mowlavi} N.,
  {Eggenberger} P.,   {Maeder} A.,  2013, \mn@doi [\aap]
  {10.1051/0004-6361/201220558}, \href
  {http://adsabs.harvard.edu/abs/2013A%26A...553A..24G} {553, A24}

\bibitem[\protect\citeauthoryear{{Gonz{\'a}lez-Fern{\'a}ndez} \&
  {Negueruela}}{{Gonz{\'a}lez-Fern{\'a}ndez} \&
  {Negueruela}}{2012}]{gonzalez12}
{Gonz{\'a}lez-Fern{\'a}ndez} C.,  {Negueruela} I.,  2012, \mn@doi [\aap]
  {10.1051/0004-6361/201118090}, \href
  {http://adsabs.harvard.edu/abs/2012A%26A...539A.100G} {539, A100}

\bibitem[\protect\citeauthoryear{{Gonz{\'a}lez-Fern{\'a}ndez}, {Dorda},
  {Negueruela}  \& {Marco}}{{Gonz{\'a}lez-Fern{\'a}ndez}
  et~al.}{2015}]{carlos15}
{Gonz{\'a}lez-Fern{\'a}ndez} C.,  {Dorda} R.,  {Negueruela} I.,   {Marco} A.,
  2015, \mn@doi [\aap] {10.1051/0004-6361/201425362}, \href
  {http://adsabs.harvard.edu/abs/2015A%26A...578A...3G} {578, A3}

\bibitem[\protect\citeauthoryear{{Gray} \& {Corbally}}{{Gray} \&
  {Corbally}}{1994}]{graco94}
{Gray} R.~O.,  {Corbally} C.~J.,  1994, \mn@doi [\aj] {10.1086/116893}, \href
  {http://adsabs.harvard.edu/abs/1994AJ....107..742G} {107, 742}

\bibitem[\protect\citeauthoryear{{Gray} \& {Corbally}}{{Gray} \&
  {Corbally}}{2009}]{graco09}
{Gray} R.~O.,  {Corbally} J. C.,  2009, {Stellar Spectral Classification}

\bibitem[\protect\citeauthoryear{{Gustafsson}, {Edvardsson}, {Eriksson},
  {J{\o}rgensen}, {Nordlund}  \& {Plez}}{{Gustafsson} et~al.}{2008}]{gus08}
{Gustafsson} B.,  {Edvardsson} B.,  {Eriksson} K.,  {J{\o}rgensen} U.~G.,
  {Nordlund} {\AA}.,   {Plez} B.,  2008, \mn@doi [\aap]
  {10.1051/0004-6361:200809724}, \href
  {http://adsabs.harvard.edu/abs/2008A%26A...486..951G} {486, 951}

\bibitem[\protect\citeauthoryear{{Harris}}{{Harris}}{1956}]{harris56}
{Harris} III D.~L.,  1956, \mn@doi [\apj] {10.1086/146174}, \href
  {http://adsabs.harvard.edu/abs/1956ApJ...123..371H} {123, 371}

\bibitem[\protect\citeauthoryear{{Howarth}, {Murray}, {Mills}  \&
  {Berry}}{{Howarth} et~al.}{2014}]{howarth14}
{Howarth} I.~D.,  {Murray} J.,  {Mills} D.,   {Berry} D.~S.,  2014, {DIPSO:
  Spectrum analysis code}, Astrophysics Source Code Library (\mn@eprint {ascl}
  {1405.016})

\bibitem[\protect\citeauthoryear{{Jeffries}}{{Jeffries}}{1997}]{jeff97}
{Jeffries} R.~D.,  1997, \mnras, \href
  {http://adsabs.harvard.edu/abs/1997MNRAS.288..585J} {288, 585}

\bibitem[\protect\citeauthoryear{{Johnson}}{{Johnson}}{1958}]{johnson58}
{Johnson} H.~L.,  1958, Lowell Observatory Bulletin, \href
  {http://adsabs.harvard.edu/abs/1958LowOB...4...37J} {4, 37}

\bibitem[\protect\citeauthoryear{{Johnson} \& {Morgan}}{{Johnson} \&
  {Morgan}}{1953}]{jm53}
{Johnson} H.~L.,  {Morgan} W.~W.,  1953, \mn@doi [\apj] {10.1086/145697}, \href
  {http://adsabs.harvard.edu/abs/1953ApJ...117..313J} {117, 313}

\bibitem[\protect\citeauthoryear{{Kharchenko}, {Piskunov}, {Schilbach},
  {R{\"o}ser}  \& {Scholz}}{{Kharchenko} et~al.}{2013}]{kharchenko13}
{Kharchenko} N.~V.,  {Piskunov} A.~E.,  {Schilbach} E.,  {R{\"o}ser} S.,
  {Scholz} R.-D.,  2013, \mn@doi [\aap] {10.1051/0004-6361/201322302}, \href
  {http://cdsads.u-strasbg.fr/abs/2013A%26A...558A..53K} {558, A53}

\bibitem[\protect\citeauthoryear{{Koornneef}}{{Koornneef}}{1983}]{koornneef}
{Koornneef} J.,  1983, \aap, \href
  {http://adsabs.harvard.edu/abs/1983A%26A...128...84K} {128, 84}

\bibitem[\protect\citeauthoryear{{Koposov} et~al.,}{{Koposov}
  et~al.}{2011}]{kopo11}
{Koposov} S.~E.,  et~al., 2011, \mn@doi [\apj] {10.1088/0004-637X/736/2/146},
  \href {http://adsabs.harvard.edu/abs/2011ApJ...736..146K} {736, 146}

\bibitem[\protect\citeauthoryear{{Kovtyukh}, {Gorlova}  \& {Belik}}{{Kovtyukh}
  et~al.}{2012}]{kovt12}
{Kovtyukh} V.~V.,  {Gorlova} N.~I.,   {Belik} S.~I.,  2012, \mn@doi [\mnras]
  {10.1111/j.1365-2966.2012.21117.x}, \href
  {http://adsabs.harvard.edu/abs/2012MNRAS.423.3268K} {423, 3268}

\bibitem[\protect\citeauthoryear{{Kroupa}}{{Kroupa}}{2001}]{kroupa01}
{Kroupa} P.,  2001, \mn@doi [\mnras] {10.1046/j.1365-8711.2001.04022.x}, \href
  {http://adsabs.harvard.edu/abs/2001MNRAS.322..231K} {322, 231}

\bibitem[\protect\citeauthoryear{{Kupka}, {Ryabchikova}, {Piskunov}, {Stempels}
   \& {Weiss}}{{Kupka} et~al.}{2000}]{kup00}
{Kupka} F.~G.,  {Ryabchikova} T.~A.,  {Piskunov} N.~E.,  {Stempels} H.~C.,
  {Weiss} W.~W.,  2000, Baltic Astronomy, \href
  {http://adsabs.harvard.edu/abs/2000BaltA...9..590K} {9, 590}

\bibitem[\protect\citeauthoryear{{Landolt}}{{Landolt}}{1992}]{landolt92}
{Landolt} A.~U.,  1992, \mn@doi [\aj] {10.1086/116242}, \href
  {http://adsabs.harvard.edu/abs/1992AJ....104..340L} {104, 340}

\bibitem[\protect\citeauthoryear{{Ma{\'{\i}}z Apell{\'a}niz}
  et~al.,}{{Ma{\'{\i}}z Apell{\'a}niz} et~al.}{2015}]{maiz15}
{Ma{\'{\i}}z Apell{\'a}niz} J.,  et~al., 2015, \mn@doi [\aap]
  {10.1051/0004-6361/201526123}, \href
  {http://adsabs.harvard.edu/abs/2015A%26A...579A.108M} {579, A108}

\bibitem[\protect\citeauthoryear{{Majaess} et~al.,}{{Majaess}
  et~al.}{2013}]{majaess13}
{Majaess} D.,  et~al., 2013, \mn@doi [\aap] {10.1051/0004-6361/201322670},
  \href {http://adsabs.harvard.edu/abs/2013A%26A...560A..22M} {560, A22}

\bibitem[\protect\citeauthoryear{{Mallik}}{{Mallik}}{1997}]{mallik97}
{Mallik} S.~V.,  1997, \mn@doi [\aaps] {10.1051/aas:1997199}, \href
  {http://adsabs.harvard.edu/abs/1997A%26AS..124..359M} {124, 359}

\bibitem[\protect\citeauthoryear{{Marco} \& {Negueruela}}{{Marco} \&
  {Negueruela}}{2017}]{marco17}
{Marco} A.,  {Negueruela} I.,  2017, \mn@doi [\mnras] {10.1093/mnras/stw2764},
  \href {http://adsabs.harvard.edu/abs/2017MNRAS.465..784M} {465, 784}

\bibitem[\protect\citeauthoryear{{Marco}, {Negueruela},
  {Gonz{\'a}lez-Fern{\'a}ndez}, {Ma{\'{\i}}z Apell{\'a}niz}, {Dorda}  \&
  {Clark}}{{Marco} et~al.}{2014}]{marco14}
{Marco} A.,  {Negueruela} I.,  {Gonz{\'a}lez-Fern{\'a}ndez} C.,  {Ma{\'{\i}}z
  Apell{\'a}niz} J.,  {Dorda} R.,   {Clark} J.~S.,  2014, \mn@doi [\aap]
  {10.1051/0004-6361/201423897}, \href
  {http://adsabs.harvard.edu/abs/2014A%26A...567A..73M} {567, A73}

\bibitem[\protect\citeauthoryear{{Martins} \& {Plez}}{{Martins} \&
  {Plez}}{2006}]{marplez06}
{Martins} F.,  {Plez} B.,  2006, \mn@doi [\aap] {10.1051/0004-6361:20065753},
  \href {http://adsabs.harvard.edu/abs/2006A%26A...457..637M} {457, 637}

\bibitem[\protect\citeauthoryear{{Mermilliod}, {Mayor}  \& {Udry}}{{Mermilliod}
  et~al.}{2008}]{mermilliod08}
{Mermilliod} J.~C.,  {Mayor} M.,   {Udry} S.,  2008, \mn@doi [\aap]
  {10.1051/0004-6361:200809664}, \href
  {http://cdsads.u-strasbg.fr/abs/2008A%26A...485..303M} {485, 303}

\bibitem[\protect\citeauthoryear{{Meynet} \& {Maeder}}{{Meynet} \&
  {Maeder}}{2000}]{mm00}
{Meynet} G.,  {Maeder} A.,  2000, \aap, \href
  {http://adsabs.harvard.edu/abs/2000A%26A...361..101M} {361, 101}

\bibitem[\protect\citeauthoryear{{Mowlavi} \& {Forestini}}{{Mowlavi} \&
  {Forestini}}{1994}]{mf94}
{Mowlavi} N.,  {Forestini} M.,  1994, \aap, \href
  {http://adsabs.harvard.edu/abs/1994A%26A...282..843M} {282, 843}

\bibitem[\protect\citeauthoryear{{Negueruela}}{{Negueruela}}{2016}]{neg16iaufm}
{Negueruela} I.,  2016, \mn@doi [IAU Focus Meeting]
  {10.1017/S1743921316005858}, \href
  {http://adsabs.harvard.edu/abs/2016IAUFM..29B.461N} {29, 461}

\bibitem[\protect\citeauthoryear{{Negueruela} \& {Marco}}{{Negueruela} \&
  {Marco}}{2012}]{negmar12}
{Negueruela} I.,  {Marco} A.,  2012, \mn@doi [\aj]
  {10.1088/0004-6256/143/2/46}, \href
  {http://adsabs.harvard.edu/abs/2012AJ....143...46N} {143, 46}

\bibitem[\protect\citeauthoryear{{Negueruela} \& {Schurch}}{{Negueruela} \&
  {Schurch}}{2007}]{ns07}
{Negueruela} I.,  {Schurch} M.~P.~E.,  2007, \mn@doi [\aap]
  {10.1051/0004-6361:20066054}, \href
  {http://adsabs.harvard.edu/abs/2007A%26A...461..631N} {461, 631}

\bibitem[\protect\citeauthoryear{{Negueruela}, {Marco},
  {Gonz{\'a}lez-Fern{\'a}ndez}, {Jim{\'e}nez-Esteban}, {Clark}, {Garcia}  \&
  {Solano}}{{Negueruela} et~al.}{2012}]{negueruela12}
{Negueruela} I.,  {Marco} A.,  {Gonz{\'a}lez-Fern{\'a}ndez} C.,
  {Jim{\'e}nez-Esteban} F.,  {Clark} J.~S.,  {Garcia} M.,   {Solano} E.,  2012,
  \mn@doi [\aap] {10.1051/0004-6361/201219540}, \href
  {http://adsabs.harvard.edu/abs/2012A%26A...547A..15N} {547, A15}

\bibitem[\protect\citeauthoryear{{Negueruela}, {Clark}, {Dorda},
  {Gonz{\'a}lez-Fern{\'a}ndez}, {Marco}  \& {Mongui{\'o}}}{{Negueruela}
  et~al.}{2016}]{neg16multi}
{Negueruela} I.,  {Clark} J.~S.,  {Dorda} R.,  {Gonz{\'a}lez-Fern{\'a}ndez} C.,
   {Marco} A.,   {Mongui{\'o}} M.,  2016, in {Skillen} I.,  {Barcells} M.,
  {Trager} S.,  eds,  Astronomical Society of the Pacific Conference Series
  Vol. 507, Multi-Object Spectroscopy in the Next Decade: Big Questions, Large
  Surveys, and Wide Fields. p.~75

\bibitem[\protect\citeauthoryear{{Netopil}, {Paunzen}  \&
  {St{\"u}tz}}{{Netopil} et~al.}{2012}]{webda}
{Netopil} M.,  {Paunzen} E.,   {St{\"u}tz} C.,  2012, \mn@doi [Astrophysics and
  Space Science Proceedings] {10.1007/978-3-642-22113-2_7}, \href
  {http://adsabs.harvard.edu/abs/2012ASSP...29...53N} {29, 53}

\bibitem[\protect\citeauthoryear{{Palacios}, {Gebran}, {Josselin}, {Martins},
  {Plez}, {Belmas}  \& {L{\`e}bre}}{{Palacios} et~al.}{2010}]{pollux}
{Palacios} A.,  {Gebran} M.,  {Josselin} E.,  {Martins} F.,  {Plez} B.,
  {Belmas} M.,   {L{\`e}bre} A.,  2010, \mn@doi [\aap]
  {10.1051/0004-6361/200913932}, \href
  {http://adsabs.harvard.edu/abs/2010A%26A...516A..13P} {516, A13}

\bibitem[\protect\citeauthoryear{{Piskunov}, {Kupka}, {Ryabchikova}, {Weiss}
  \& {Jeffery}}{{Piskunov} et~al.}{1995}]{pis95}
{Piskunov} N.~E.,  {Kupka} F.,  {Ryabchikova} T.~A.,  {Weiss} W.~W.,
  {Jeffery} C.~S.,  1995, \aaps, \href
  {http://adsabs.harvard.edu/abs/1995A%26AS..112..525P} {112, 525}

\bibitem[\protect\citeauthoryear{{Poelarends}, {Herwig}, {Langer}  \&
  {Heger}}{{Poelarends} et~al.}{2008}]{poelarends08}
{Poelarends} A.~J.~T.,  {Herwig} F.,  {Langer} N.,   {Heger} A.,  2008, \mn@doi
  [\apj] {10.1086/520872}, \href
  {http://adsabs.harvard.edu/abs/2008ApJ...675..614P} {675, 614}

\bibitem[\protect\citeauthoryear{{Raskin } \& {Van Winckel}}{{Raskin } \& {Van
  Winckel}}{2014}]{raskin14}
{Raskin } G.,  {Van Winckel} H.,  2014, \mn@doi [Astronomische Nachrichten]
  {10.1002/asna.201312009}, \href
  {http://adsabs.harvard.edu/abs/2014AN....335...32R} {335, 32}

\bibitem[\protect\citeauthoryear{{Reid} et~al.,}{{Reid} et~al.}{2014}]{reid14}
{Reid} M.~J.,  et~al., 2014, \mn@doi [\apj] {10.1088/0004-637X/783/2/130},
  \href {http://adsabs.harvard.edu/abs/2014ApJ...783..130R} {783, 130}

\bibitem[\protect\citeauthoryear{{Rieke} \& {Lebofsky}}{{Rieke} \&
  {Lebofsky}}{1985}]{riekes}
{Rieke} G.~H.,  {Lebofsky} M.~J.,  1985, \mn@doi [\apj] {10.1086/162827}, \href
  {http://adsabs.harvard.edu/abs/1985ApJ...288..618R} {288, 618}

\bibitem[\protect\citeauthoryear{{Ryabchikova}, {Piskunov}, {Kurucz},
  {Stempels}, {Heiter}, {Pakhomov}  \& {Barklem}}{{Ryabchikova}
  et~al.}{2015}]{rya15}
{Ryabchikova} T.,  {Piskunov} N.,  {Kurucz} R.~L.,  {Stempels} H.~C.,  {Heiter}
  U.,  {Pakhomov} Y.,   {Barklem} P.~S.,  2015, \mn@doi [\physscr]
  {10.1088/0031-8949/90/5/054005}, \href
  {http://adsabs.harvard.edu/abs/2015PhyS...90e4005R} {90, 054005}

\bibitem[\protect\citeauthoryear{{Salasnich}, {Bressan}  \&
  {Chiosi}}{{Salasnich} et~al.}{1999}]{salas99}
{Salasnich} B.,  {Bressan} A.,   {Chiosi} C.,  1999, \aap, \href
  {http://adsabs.harvard.edu/abs/1999A%26A...342..131S} {342, 131}

\bibitem[\protect\citeauthoryear{{Schaller}, {Schaerer}, {Meynet}  \&
  {Maeder}}{{Schaller} et~al.}{1992}]{schaller92}
{Schaller} G.,  {Schaerer} D.,  {Meynet} G.,   {Maeder} A.,  1992, \aaps, \href
  {http://cdsads.u-strasbg.fr/abs/1992A%26AS...96..269S} {96, 269}

\bibitem[\protect\citeauthoryear{{Shortridge}, {Meyerdierks}, {Currie},
  {Davenhall}, {Jenness}  \& {Clayton}}{{Shortridge} et~al.}{2014}]{shortridge}
{Shortridge} K.,  {Meyerdierks} H.,  {Currie} M.~J.,  {Davenhall} C.,
  {Jenness} T.,   {Clayton} M.,  2014, {Starlink Figaro: Starlink version of
  the Figaro data reduction software package}, Astrophysics Source Code Library
  (\mn@eprint {ascl} {1411.022})

\bibitem[\protect\citeauthoryear{{Skrutskie} et~al.,}{{Skrutskie}
  et~al.}{2006}]{skru06}
{Skrutskie} M.~F.,  et~al., 2006, \mn@doi [\aj] {10.1086/498708}, \href
  {http://adsabs.harvard.edu/abs/2006AJ....131.1163S} {131, 1163}

\bibitem[\protect\citeauthoryear{{Stetson}}{{Stetson}}{1987}]{stetson87}
{Stetson} P.~B.,  1987, \mn@doi [\pasp] {10.1086/131977}, \href
  {http://adsabs.harvard.edu/abs/1987PASP...99..191S} {99, 191}

\bibitem[\protect\citeauthoryear{{Subramaniam}, {Carraro}  \&
  {Janes}}{{Subramaniam} et~al.}{2010}]{subram10}
{Subramaniam} A.,  {Carraro} G.,   {Janes} K.~A.,  2010, \mn@doi [\mnras]
  {10.1111/j.1365-2966.2010.16345.x}, \href
  {http://cdsads.u-strasbg.fr/abs/2010MNRAS.404.1385S} {404, 1385}

\bibitem[\protect\citeauthoryear{{Tabernero}, {Dorda}, {Negueruela}  \&
  {Gonz{\'a}lez-Fern{\'a}ndez}}{{Tabernero} et~al.}{2018}]{tab18}
{Tabernero} H.~M.,  {Dorda} R.,  {Negueruela} I.,
  {Gonz{\'a}lez-Fern{\'a}ndez} C.,  2018, \mn@doi [\mnras]
  {10.1093/mnras/sty399}, \href
  {http://adsabs.harvard.edu/abs/2018MNRAS.tmp..381T} {}

\bibitem[\protect\citeauthoryear{{Tadross}}{{Tadross}}{2008}]{tadross}
{Tadross} A.~L.,  2008, \mn@doi [\mnras] {10.1111/j.1365-2966.2008.13554.x},
  \href {http://cdsads.u-strasbg.fr/abs/2008MNRAS.389..285T} {389, 285}

\bibitem[\protect\citeauthoryear{{Turner}}{{Turner}}{1980}]{turner80}
{Turner} D.~G.,  1980, \mn@doi [\apj] {10.1086/158216}, \href
  {http://adsabs.harvard.edu/abs/1980ApJ...240..137T} {240, 137}

\bibitem[\protect\citeauthoryear{{Weidemann}}{{Weidemann}}{2000}]{weide00}
{Weidemann} V.,  2000, \aap, \href
  {http://adsabs.harvard.edu/abs/2000A%26A...363..647W} {363, 647}

\bibitem[\protect\citeauthoryear{{Winkler}}{{Winkler}}{1997}]{winkler97}
{Winkler} H.,  1997, \mnras, \href
  {http://adsabs.harvard.edu/abs/1997MNRAS.287..481W} {287, 481}

\bibitem[\protect\citeauthoryear{{Xu} et~al.,}{{Xu} et~al.}{2013}]{xu13}
{Xu} Y.,  et~al., 2013, \mn@doi [\apj] {10.1088/0004-637X/769/1/15}, \href
  {http://adsabs.harvard.edu/abs/2013ApJ...769...15X} {769, 15}

\bibitem[\protect\citeauthoryear{{van Leeuwen}}{{van
  Leeuwen}}{2009}]{leeuwen09}
{van Leeuwen} F.,  2009, \mn@doi [\aap] {10.1051/0004-6361/200811382}, \href
  {http://adsabs.harvard.edu/abs/2009A%26A...497..209V} {497, 209}

\makeatother
\end{thebibliography}




\appendix
\section{Synthetic clusters}
\label{synth}

One hundred synthetic clusters were kindly generated by Dr. Sylvia Ekstr\"{o}m by using the interactive tools made available by the Geneva group \citep{georgy17}. Each cluster contains 200 stars with masses ranging between $1.7\:\mathrm{M}_{\sun}$ and the highest mass in the cluster. This highest mass depends on the stochastic sampling of the IMF and the initial rotational velocity distribution; it ranges between $6.5\:\mathrm{M}_{\sun}$ and $7.1\:\mathrm{M}_{\sun}$, with a typical (both median and mode) value of $6.9\:\mathrm{M}_{\sun}$. The 200 stars (a fraction of which are binaries) have a mass ranging from~635 to~$760\:\mathrm{M}_{\sun}$, with an average of $695\:\mathrm{M}_{\sun}$. For a standard IMF, this translates into an initial total mass of $\sim3\,000\:\mathrm{M}_{\sun}$ for each cluster.

To analyse the synthetic CMDs, we defined four classes of post-MS stars:
\begin{itemize}
\item \textbf{Blue giants: }These are stars just above the turnoff, but probably still burning H in their cores. We select objects with $(B-V)\leq-0.05$ and $-3\leq M_{V}\leq-4.5$, typical of B2\,--\,B8 giants. 
\item \textbf{Blue supergiants: }We select objects with $(B-V)\leq-0.05$ and $M_{V}\leq-4.5$ (B-type supergiants) and objects with $(B-V)>-0.05$, $(U-B)<-0.05$ and $M_{V}\leq-4.0$ (A-type supergiants).
\item \textbf{Yellow supergiants: }We select objects with $(U-B)\geq-0.05$, $(B-V)<1.00$ and $M_{V}<-3.0$. The first condition implies a division between blue and yellow supergiants around spectral type A8; the second condition divides yellow from red supergiants at G5.
\item \textbf{Red supergiants: }These are selected as having $(B-V)\geq1.00$, $M_{V}\leq-2.5$.
\end{itemize}

%
   \begin{figure}
   \centering
 \resizebox{\columnwidth}{!}{\includegraphics[clip]{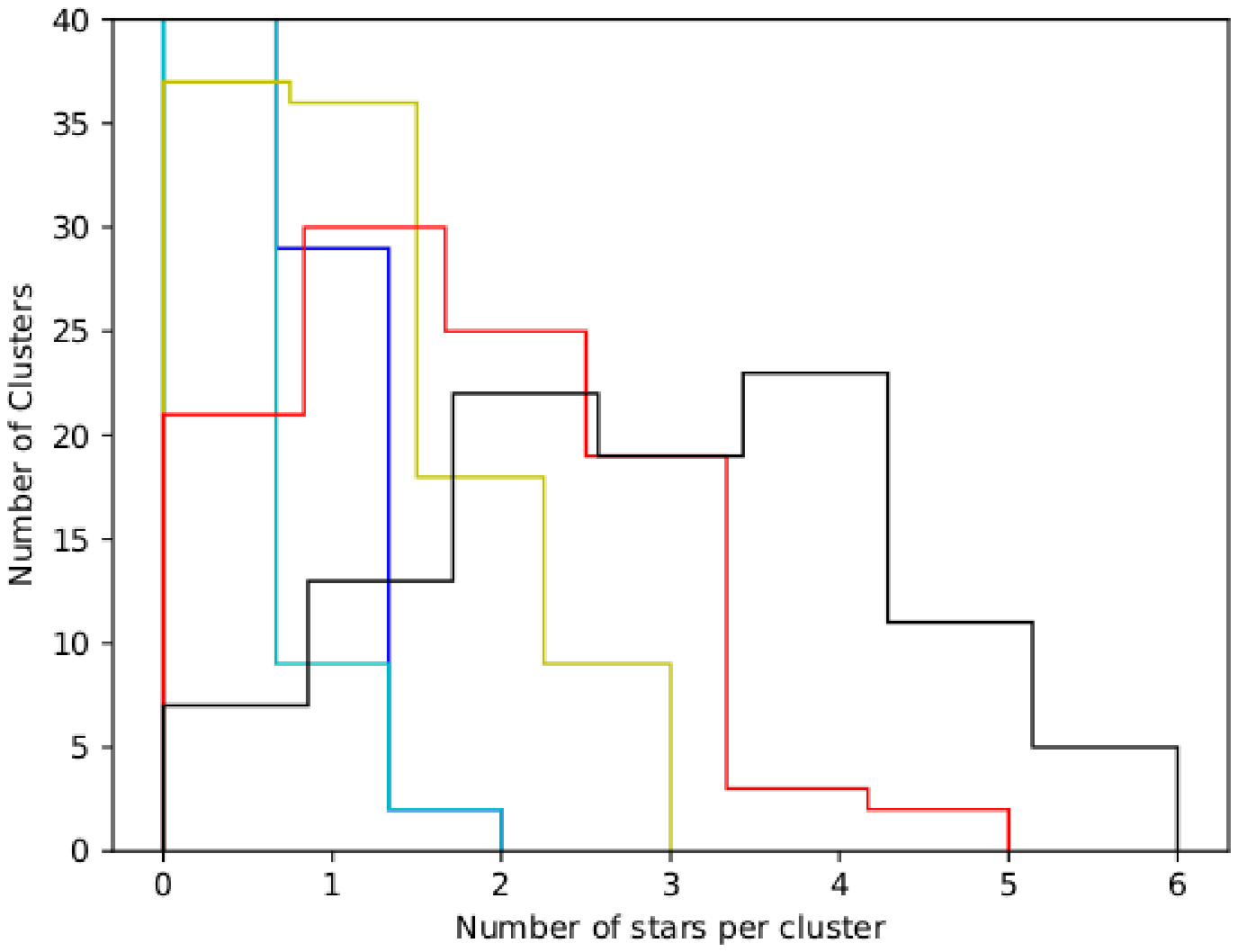}}
       \caption{Distribution of evolved stars in the synthetic clusters. The plot represents the number of clusters in the sample containing a given number of evolved stars: blue giants (cyan); blue supergiants (blue); yellow supergiants (yellow); red supergiants (red); total number of supergiants (black). The vertical scale has been cut for clarity. The number of clusters containing zero blue giants is 89, while the number of clusters containing zero blue supergiants is 69. 
         \label{synth_byclass}}
   \end{figure}
%

The total number of evolved stars found in the 100 synthetic clusters is 304, i.e.\ a cluster of $\sim3\,000$M$_{\sun}$ is typically expected to have 3 evolved stars. There are only 11 clusters with blue giants, with a total of 13 blue giants. Among the supergiants, the numbers are 33/99/159 for blue/yellow/red (as a fraction of the total number, 0.11/0.34/0.55 in good agreement with the 20\,000 star synthetic cluster discussed in Sect.~\ref{params}). The distribution of evolved stars in the clusters is shown in Fig.~\ref{synth_byclass}.

\section{Photometric data}
\label{data}
\onecolumn

\begin{landscape}
\begin{longtable}[h]{ c c c | c c c c c c c | c c c c c c c c }
\caption{Coordinates, $UBV$ photometry, 2MASS identification and 2MASS $JHK_{\textrm{S}}$ photometry for all stars in the field of Berkeley~51. Errors in the $UBV$ photometry represent the standard deviation of $n$ measurements or the photometric error when $n=1$. Errors in 2MASS data are as given by the catalogue. The flags identify the quality of 2MASS photometry, with U indicating upper limits. \textit{Only the first ten entrances are shown. The whole table will appear as material on-line.}\label{allphot}}\\
\hline\hline
\noalign{\smallskip}
RA (J2000 )& Dec (J2000)& Name &$V$&$\sigma_V$&$(B-V)$ & $\sigma_{(B-V)}$ &$(U-B)$ & $\sigma_{(U-B)}$ & $n$ & 2MASS ID &$J$&$E_J$&$H$&$E_H$&$K_{\textrm{S}}$& $E_{K_{\textrm{S}}}$& Flag\\
\noalign{\smallskip}
\hline
\noalign{\smallskip}
\endhead
20:11:48.99&+34:21:14.5&1&16.874&0.006&1.067&0.008&0.628&0.012&4&20114899+3421145&14.208&0.027&13.848&0.033&13.584&0.041&AAA\\
20:11:38.99&+34:21:26.5&2&16.712&0.005&1.344&0.008&0.518&0.011&4&20113899+3421265&14.002&0.028&13.509&0.027&13.327&0.036&AAA\\
20:11:51.53&+34:21:27.2&3&17.312&0.007&1.157&0.010&0.439&0.016&4&20115153+3421272&14.756&0.037&14.345&0.054&14.093&0.058&AAA\\
20:12:04.48&+34:21:30.7&4&19.118&0.020&1.515&0.033&0.886&0.099&1&20120448+3421307&15.166&0.050&14.006&--&13.494&--&AUU\\
20:11:57.81&+34:21:33.6&6&18.465&0.013&1.285&0.021&0.768&0.047&2&20115781+3421336&15.821&0.070&15.234&0.091&15.341&0.163&AAC\\
20:11:44.10&+34:21:36.2&7&17.646&0.008&1.408&0.014&0.876&0.029&4&20114410+3421362&14.845&0.037&14.302&0.048&14.131&0.063&AAA\\
20:12:03.67&+34:21:33.9&8&14.929&0.002&0.895&0.002&0.619&0.002&4&20120367+3421339&12.768&0.023&12.499&0.023&12.321&0.023&AAA\\
20:11:38.67&+34:21:37.4&9&17.050&0.007&1.499&0.011&0.707&0.018&3&20113867+3421374&14.037&0.025&13.540&0.026&13.333&0.035&AAA\\
20:11:48.97&+34:21:38.5&10&16.925&0.006&1.363&0.009&1.444&0.022&4&20114897+3421385&14.038&0.026&13.394&0.029&13.185&0.035&AAA\\
20:11:58.45&+34:21:50.4&12&18.311&0.013&1.394&0.020&0.802&0.044&2&20115845+3421504&15.061&0.041&14.704&0.055&14.367&0.062&AAA\\
\noalign{\smallskip}
\hline
\end{longtable}
\end{landscape}


\bsp	
\label{lastpage}
\end{document}